\documentclass[manuscript,screen, nonacm]{acmart}
\AtBeginDocument{%
  \providecommand\BibTeX{{%
    \normalfont B\kern-0.5em{\scshape i\kern-0.25em b}\kern-0.8em\TeX}}}

\setcopyright{acmcopyright}
\copyrightyear{2018}
\acmYear{2018}
\acmDOI{XXXXXXX.XXXXXXX}

\acmConference[CSCW]{Make sure to enter the correct
  conference title from your rights confirmation emai}{ 
  2025}{  }
\acmBooktitle{Woodstock '18: 
 June 03--05, 2025, Woodstock, NY} 
\acmPrice{15.00}
\acmISBN{978-1-4503-XXXX-X/18/06}

\usepackage{tikz}
\usepackage{amsmath}
\usepackage{filecontents}
\usepackage{titlesec}
\titleclass{\subsubsubsection}{straight}[\subsubsection]
\newcounter{subsubsubsection}[subsubsection]
\renewcommand\thesubsubsubsection{\thesubsubsection.\arabic{subsubsubsection}}
\titleformat{\subsubsubsection}
  {\normalfont\normalsize\itshape}{\thesubsubsubsection}{1em}{}
\titlespacing*{\subsubsubsection}
  {0pt}{1.25ex plus 1ex minus .2ex}{0.5ex plus .2ex}
\usepackage{tabularx}
\usepackage{booktabs}
\usepackage{graphicx}
\usepackage{hyperref}
\usepackage{hhline}
\usepackage{subcaption}
\usepackage{enumitem}

\usepackage{float} 
\usepackage{soul,color}
\usepackage[utf8]{inputenc}
\usepackage{longtable}
\usepackage[flushleft]{threeparttable}
\usepackage{chngcntr}
\usepackage{xcolor}
\definecolor{editCol}{rgb}{0.0, 0.0, 0.0}
\newcommand{\edit}[1]{{\textcolor{editCol}{#1}}}
\usepackage{pbox}
\usepackage{multirow}

\usepackage{array}
\usepackage{makecell}
\usepackage{afterpage} 
\usepackage[title]{appendix}
\usepackage{subcaption,graphicx}
\usepackage{ragged2e}

\usepackage{tabularx} 
\usepackage{multirow} 

\usepackage{graphicx}

\begin{document}

\title{Unfiltered: How Teens Engage in Body Image and Shaming Discussions via Instagram Direct Messages (DMs)}

\author{Abdulmalik Alluhidan}
\email{abdulmalik.s.alluhidan@vanderbilt.edu}
\orcid{0000-0001-8948-3923}
\affiliation{
  \institution{Vanderbilt University}
  \city{Nashville}
  \state{Tennessee}
  \country{USA}
}
\author{Jinkyung Katie Park}
\email{Jinkyup@clemson.edu}
\orcid{0000-0002-0804-832X}
\affiliation{%
  \institution{Clemson University}
  \city{Clemson}
  \state{South Carolina}
  \postcode{29634}
  \country{USA}
}

\author{Mamtaj Akter}
\email{Mamtaj.Akter@nyit.edu}
\orcid{0000-0002-5692-9252}
\affiliation{%
  \institution{New York Institute of Technology}
  \city{New York}
  \state{New York}
  \postcode{10023}
  \country{USA}
}

\author{Rachel Rodgers}
\email{r.rodgers@northeastern.edu}
\orcid{0000-0002-2192-9332}
\affiliation{%
  \institution{Northeastern University}
  \city{Boston}
  \state{Massachusetts}
  \postcode{02115}
  \country{USA}
}
\author{Afsaneh Razi}
\email{afsaneh.razi@drexel.edu}
 \orcid{0000-0001-5829-8004}
\affiliation{%
  \institution{Drexel University}
   \streetaddress{3675 Market St 10th floor}
  \city{Philadelphia}
   \state{Pennsylvania}
   \country{U.S.A}
   \postcode{19104}
}
\author{Pamela J. Wisniewski}
\email{Pamwis@stirlab.org}
\orcid{0000-0002-6223-1029}
\affiliation{%
  \institution{Socio-Technical Interaction Research Lab}
   \country{USA}
}

\renewcommand{\shortauthors}{Alluhidan et al.}

\begin{abstract}
We analyzed 1,596 sub-conversations within 451 direct message (DM) conversations from 67 teens (ages 13-17) who engaged in private discussions about body image on Instagram. Our findings show that teens often receive support when sharing struggles with negative body image, participate in criticism when engaging in body-shaming, and are met with appreciation when promoting positive body image. Additionally, these types of disclosures and responses varied based on whether the conversations were one-on-one or group-based. We found that sharing struggles and receiving support most often occurred in one-on-one conversations, while body shaming and negative interactions often occurred in group settings. A key insight of the study is that private social media settings can significantly influence how teens discuss and respond to body image. 
Based on these findings, we suggest design guidelines for social media platforms that could promote positive interactions around body image, ultimately creating a healthier and more supportive online environment for teens dealing with body image concerns.
\newline
\textbf{Content Warning:} This paper discusses sensitive topics related to body image and shaming. It also includes explicit language. Reader discretion is advised.

\end{abstract}







\maketitle

\section{Introduction}

Body image is a critical aspect of adolescent development, significantly impacting mental health and self-esteem \cite{ajmal2019impact, choukas2022perfect}. In 2022, a nationally representative report \edit{in the United States} showed that 73\% of teenage girls and 69\% of teenage boys had expressed significant self-consciousness regarding their physical appearance \cite{Mostafavi_2022}.
With their visually centered design and social features, social media platforms such as Instagram have become central to how young people engage with one another about appearance~\cite{alluhidan2024teen, devito2017platforms}. Such public social media platforms have been reported to exacerbate body image concerns by showcasing idealized images, leading to increased body dissatisfaction through heightened social comparisons~\cite{pedalino2022instagram, jung2022social}.
In addition, within these environments, peer influences have been shown to play a crucial role in either exacerbating their body image concerns or helping to alleviate them~\cite{mills2019impact, ambwani2017challenging}.  
Online peer support is increasingly gaining traction within the SIGCHI community~\cite{chen2021scaffolding, huh2023help,lambton2021blending}. 
However, earlier studies on body image support-seeking have mostly focused on exploring online eating disorder communities \cite{nova2022cultivating, pater2019notjustgirls, devakumar2021review}, offering limited insight into how teens engage in broader discussions about body image. Therefore, examining how teens disclose body image concerns and the responses they receive from peers comprehensively is essential for gaining a deeper understanding of the support dynamics that exist within online interactions.

In addition, the distinction between teens' engagement on public versus private social media platforms \cite{rubio2022youths, deogracias2015danah, marwick2014networked} further motivates investigation of teens' peer support in private online settings; notably, young individuals' self-expression in private settings often includes more personal and sensitive disclosures compared to their more reserved public posts \cite{rubio2022youths}. This discrepancy suggests that private conversations may offer deeper insights into adolescents' body image concerns and the support dynamics that unfold away from the public eye.
Particularly, prior research \cite{alluhidan2024teen} has shown that teens frequently mention Instagram when seeking support for body image issues they encounter on social media given its visually centered design and interactive features. 
This finding further motivates the need to understand private conversations on how teens discuss body-image issues with peers through private online chats on Instagram. 
In this work, we aim to understand not only the types of body image disclosures teens share but also how others respond to these discussions in private conversations to provide implications for designing effective support.
It is also important to explore the dynamics of these conversations, particularly whether differences arise in dyadic (one-on-one) versus group settings. The context of these interactions plays a crucial role in shaping both the nature of the disclosures and the responses received \cite{kruger2019individual, mills2019impact}. As such, we pose the following high-level research questions:

\begin{itemize}
\item \textbf{RQ1:} \textit{What types of conversations do teens have regarding body image through private social media channels?} 
\item \textbf{RQ2:} \textit{How do their conversation partners respond to the topics raised?} 
\item \textbf{RQ3:} \textit{Are there unique differences in the types of topics and responses based on whether the conversations are dyadic or group-based?} 
\end{itemize}

To address these research questions, we used a dataset collected through the Instagram Data Donation (IGDD) study 
\cite{razi2022instagram}, which included direct and group private messages shared by youth aged 13-21. \edit{We employed a mixed methods approach, integrating qualitative thematic analysis with quantitative statistical testing.} We conducted a qualitative analysis of 1,596 sub-conversations within 451 private message conversations from 67 participants aged 13-17. Using thematic qualitative analysis \cite{braun2012thematic}, we identified key themes from the data to answer our RQ1 and RQ2. To answer our RQ3, we utilized our structured coding process to conduct chi-square tests, allowing us to identify statistical differences in themes based on the type of conversation (one-on-one vs. group).

Overall, our findings revealed that conversations on body image primarily centered on struggles with negative body image. Teens frequently judged themselves and attempted to alter their appearance. In private dyadic conversations, they were much more likely to open up about these struggles and receive support. Peers typically provided support to these disclosures by offering validation and sharing their own similar experiences. On the other hand, such disclosures and supportive responses were less common in group settings. We also observed that discussions about body image often involved body shaming, with exchanges ranging from serious to humorous tones. Body shaming occurred more frequently in group settings, while dyadic conversations were less likely to involve such criticism. Responses to body shaming were generally negative, as peers often criticized each other in return; however, on a few occasions, they defended those who were targeted. On a more positive note, we found instances where conversations promoted positive body image. In these cases, teens uplifted others and encouraged self-acceptance. Such positive discussions were usually well-received, with peers appreciating the sentiments and offering validation in return. 

Based on these findings, we make these novel contributions to the CSCW and SIGCHI community:
 \begin{itemize}
\item Advance the understanding of teens' disclosures about their body image in private online conversations by examining how sensitive concerns are shared in closed social settings.
\item Provide insights into the types of responses teens receive to their body image disclosures, illustrating how peers offer support or criticism in private online interactions.
\item Demonstrate the unique differences in topics and responses based on whether the conversations are dyadic or group-based, revealing how the conversation setting impacts the nature of body image discussions and peer support among teens.
\item Offer evidence-based design guidelines that encourage positive peer support around body image, providing researchers and designers with actionable insights to proactively create interventions that foster healthier and more cooperative social environments.

 \end{itemize}


\section{Background}
In this section, we review the relevant literature on how youth seeks support for body image concerns and explore the nuances of examining peer conversations about body image.

\subsection{Youth Body Image in the Age of Social Media} As an individual's perceptions, thoughts, attitudes, and feelings about their own physical appearance, body image is a critical aspect of adolescent development, significantly impacting mental health and self-esteem \cite{ajmal2019impact, choukas2022perfect}. 
Past studies have extensively examined the connection between the use of social media and concerns related to body image, describing body image as an individual's perceptions, thoughts, and feelings regarding their appearance \cite{grogan2021body}. Specifically, findings consistently showed that exposure to idealized and digitally modified images on social media platforms can adversely impact youth body image \cite{vandenbosch2022social, saiphoo2019meta, faelens2021relationship}. 
A particular focus has been placed on the increase of idealized imagery on social media, 
which have been identified as harmful to mental health in various aspects, including body dissatisfaction \cite{anixiadis2019effects, brown2020picture}, facial dissatisfaction \cite{sampson2020effect, fardouly2019impact}, and self-objectification, viewing and valuing the body from an external viewpoint  \cite{Cohen2019BoPo}.
The effects of social media on body image are complex and can vary among individuals \cite{mahon2021processing}.
Within the HCI research, the study of user behaviors and experiences on social media has been conducted through the lens of "affordances," which describes the interactive relationship between users and social media platforms \cite{faraj2012materiality}. This principle has served as a fundamental concept in exploring how social media's characteristics can either facilitate or limit user behavior on these platforms \cite{park2023affordances, ronzhyn2023defining}. 
For instance, 
research by Fatt and Fardouly \cite{fatt2023digital} indicates that adolescents who often receive positive feedback on their appearance on social media, along with those who place significant value on features such as the number of followers and ''likes,'' face an increased risk of developing body image concerns. 
Furthermore, 
prior studies have demonstrated that social media users often use content persistence affordances, \edit{the ability of a platform to keep content accessible over time unless manually deleted (e.g., Instagram posts)} \cite{ronzhyn2023defining}, to project an idealized self-image. Meanwhile, they rely on ephemeral affordances, \edit{the capacity of a platform to make content temporarily available (e.g., Instagram Stories)} \cite{kim2023social}, to showcase more authentic and casual expressions \cite{chiu2021last}. This body of research collectively suggests social media affordances, from the type of comments received to the perceived value of online affirmations and the choice of social media affordances, significantly shape users' perceptions of their bodies. Indeed, recent research showed that the issue of body image was frequently mentioned by youth when referring to Instagram \cite{alluhidan2024teen}, one of the popular platforms with visually centered design and interactive affordances, which signifies the need to understand how youth discuss body image issues with peers on Instagram. 
In this work, we examined youth's private conversations on body-image issues with peers on Instagram to provide implications for designing social media environments to support their positive body image.

\subsection{Youth Body Image and Online Peer Support}

Online peer support is gaining importance within the SIGCHI community~\cite{ding2023infrastructural, chen2021scaffolding, huh2023help, oguine2024internet}. Previous research shows that youth go online to seek social and peer support for sensitive matters like self-harm \cite{thorn2023motivations}, romantic relationships \cite{kim2017romantic}, and sexual experiences \cite{razi2020let}. Online peer conversations provide emotional support and validation \cite{syed2024machine}, helping users cope with negative emotions and mental health challenges. Peer-support platforms allow individuals to share experiences and foster connection~\cite{eagle2023you}. Studies have also explored peer support on body image, with a focus on online eating disorder communities \cite{nova2022cultivating, pater2019notjustgirls, fettach2019pro}. For instance, eating disorder recovery communities provide a space for discussing treatment options and challenges, creating a supportive environment for those seeking help~\cite{fettach2019pro}.
Despite benefits, such interactions also carry risks. 
Unmoderated environments sometimes reinforce negative behaviors when harmful habits are validated~\cite{kruzan2021investigating}. For instance, Twitter eating disorder communities often feature harmful content spread by influencers\cite{nova2022cultivating}, while Instagram users use banned tags to continue sharing pro-eating disorder content \cite{pater2019exploring}. 
Therefore, understanding the dual nature of peer interactions on body image disclosures is essential for designing effective support systems for youth. This work examines youth's support-seeking conversations and responses to identify ways to promote positive body image and mitigate risks of peer-influenced negativity. 

Meanwhile, the dynamics of peer support on body image can vary depending on different online settings. For instance, discussions in public forums are fundamentally distinct from those in private settings~\cite{mcgregor201973}, as they offer a secure and trusted space for sharing personal experiences and emotions \cite{gibson2020young}. 
While prior research has illustrated how adolescents seek online support, most studies have relied on publicly available social media data, such as YouTube comments ~\cite{naslund2014naturally} or Reddit posts~\cite{de2014mental}.
Previous research has shown that private online channels provide youth with supportive environments for sharing both positive and negative self-disclosures \cite{luo2020self,huh2023help}. Yet, our understanding of body image support in private conversations among young people remains limited, underscoring the need to explore these less visible, private channels further.
We seek to explore these less visible private channels further, aiming to understand how youth express and navigate their concerns regarding body image in settings where they feel safer and less exposed.
In addition, empirical studies suggest that individuals disclose less personal information in larger groups than in dyadic conversations, especially when it comes to highly intimate content \cite{solano1985two, taylor1979sharing, drag1970experimenter}.
Because such content is more emotionally charged and involves greater risks of judgment or misunderstanding by others, individuals are less likely to disclose it in a group setting \cite{fleming1991mixed, cooney2020many}. On the other hand, dyadic interactions might offer a setting where teens feel more at ease discussing personal body image concerns without fear of widespread judgment, allowing for deeper insights into individual experiences. 
As such, examining dyadic and group private conversations surrounding body image is crucial as they 
could inform different implications for supporting youth with positive body image. Yet, whether there is a difference between dyadic and group private conversation about body image exists, and if it does, how they differ has been under-studied.  By comparing these two contexts—dyadic and group private conversation, our study aims to understand the different ways peer interactions shape adolescents's conversation on body image.
\section{Methods}
In this section, we outline the dataset we analyzed, detail our scoping process, and explain our qualitative analysis. 

\subsection{Instagram Dataset and Ethical Considerations}
We used data that was originally collected as part of the Instagram Data Donation project (IGDD) 
\cite{razi2022instagram}. The data collection spanned from 2020 to the middle of 2022 and contained over seven million messages in 32,055 conversations from 195 youth Instagram accounts. 
Participants were all between the ages of 13-21 and possessed the account for at least 3 months when they were teens between the ages of 13 and 17 and had DMs with at least 15 people. 
All participants were English-speaking U.S. residents. With the necessary parental consent for minors and their assent, they were asked to download their Instagram data and upload it to a secure, web-based platform for data collection.

In terms of ethical considerations, we obtained Institutional Review Board (IRB) approval from the first author’s institution to use the dataset. Every member of the research team completed IRB CITI training \cite{hadden2018readability}, which includes courses on research ethics and the protection of human subjects. Additionally, they underwent Protection of Minors (POM) Training and a thorough background check before accessing the dataset. 
Also, to protect participants’ privacy, we handled the messages of participants and their conversation partners with the utmost care. We removed any personally identifiable information from quotations presented in this paper and sometimes paraphrased them as an extra precaution to ensure confidentiality. Additionally, we performed our qualitative analyses on university-approved and secured shared storage, preventing team members from uploading any data to the cloud or their individual computers.

 \subsection{Data Scoping and Relevancy Coding}
Our primary goal was to examine the ways teens (ages 13-17) talk about body image with their peers and what types of responses they receive via social media private messages. \edit{The rationale for selecting this specific age group lies in its unique position: while teens are still legally minors, they are not afforded the same protections as younger children under laws like the U.S. Children’s Online Privacy Protection Act (COPPA) \cite{federal2013children}. To comply with COPPA regulations, social media platforms frequently enforce restrictions, including blocking adults from sending private messages to minors they are not connected with and employing sensitivity filters or enhanced parental controls \cite{instagram_safety_2021, instagram_sensitive_content_2021}. However, these measures are limited to safeguarding children under 13, leaving those aged 13-17 without the same level of protection online. This gap is particularly concerning as adolescents are uniquely vulnerable to body image concerns due to their developmental characteristics. Specifically, teens in this age group typically experience puberty-driven physical changes \cite{dion2015development}, undergo identity formation \cite{perez2024social}, and develop the capacity for self-reflection \cite{toenders2024developing}, which makes targeting this group is especially important.}

To accomplish our goal, we first ran a query on the larger dataset to search for conversations with messages that contained specific terms associated with body image and body shaming. We leveraged previous literature \cite{corradini2023dark} to search for common terms on social media (e.g., body, ugly, fat, skinny).  This initial query retrieved 
5,526 conversations from 193 unique users. Initial examination of messages in these conversations revealed that some keywords returned no hits, hence, they were immediately excluded. Also, we found that some keywords produced a significantly higher number of irrelevant messages than others. Therefore, we decided to examine all keywords individually. To determine relevancy, we analyzed the first 100 conversations for each keyword. This analysis enabled the exclusion of more keywords that yielded no relevant hits (see Table \ref{Keywords}). 
Following this process, we performed a refined search with the remaining keywords, resulting in 927 conversations from 98 distinct users, all selected for relevancy coding.

Conversations were considered relevant if they included references to body image disclosures. We also included conversations where the dialogue incorporated the use of body shaming language, including instances where such language was not used in a serious manner.
Conversations that did not meet these criteria were excluded. Some of the conversations were coded as irrelevant due to the alternative use of some of the keywords we used in our search. For instance, some messages included the use of words such as “weight” in contexts unrelated to body image, such as in describing alcohol ("\textit{e.g., light weight}") or “fat” in other conversational expressions ("\textit{e.g., Now that’s a big fat mood}"). 
\begin{table}[h!]
\centering
\caption{Keywords for the Search Query}
\label{Keywords}
\begin{tabular}{|p{3cm}|p{11cm}|} 
\hline 
\textbf{Search Results} & \textbf{Keywords} \\ 
\hline 
Relevant & body, ugly, fat, skinny, weight, beauty, unhealthy, appearance, insecure, criticize, disgusting  \\ 
\hline 
No hits & stigma, discrimination, mock, flaw, prejudice, praise, tease, shame, ridicule \\ 
\hline 
Coded as irrelevant & hate, health, insult, looks, perfect, stereotype, obsessed, opinion, judgment \\ 
\hline 
\end{tabular}
\end{table}

In our study, 
a conversation is defined as the entire exchange of messages between participants, whether in one-on-one or group chats. These conversations can range from a few minutes to several days or even years. The average duration of these conversations was 226 days, 10 hours, and 49 minutes, with a standard deviation of 308 days, 13 hours, and 58 minutes.
A conversation does not need to follow a coherent thread and can include multiple discussion topics. Each conversation was divided into sub-conversations to facilitate data analysis. A 'sub-conversation' is a specific segment where a body image or body shaming discussion begins and ends, identified by four components: 1) initiation of the discussion, 2) the context of the disclosure, 3) the responses of the participants and their conversation partners, and 4) the conclusion, often marked by a change in subject. Often, a single conversation contained multiple sub-conversations, reflecting a history of these discussions over time. Following this extensive process of relevancy coding, our final dataset included 1,596 sub-conversations within 451 unique conversations from 67 unique users, which served as the basis for our qualitative analysis. 
Additionally, to better understand the context of these interactions, we analyzed all media files exchanged in the sub-conversations. A total of 156 media files were scoped for relevancy, and 96 relevant files were identified. These consisted of 66\% pictures (n = 105), 17\% screenshots (n = 26), 6\% videos (n = 8), 5\% memes (n = 8), 4\% drawings (n = 7), and 3\% voice notes (n = 4).


\begin{table}[!ht]
    \centering
    \caption{Codebook for RQ1 and RQ2}
    
    \footnotesize
    \begin{tabular}{|p{2.5cm}|p{3cm}|p{4cm}|p{3.5cm}|}
    \hline
    \textbf{Themes (RQ1)} & \textbf{Codes} & \textbf{Illustrative Quotations} & \textbf{Responses (RQ2)} \\ \hline
    
    \multirow{6}{2.6cm}{\textbf{Teens disclosed their struggles with negative body image} \newline (68\%, n=1,094)} 
    & \textbf{Judged themselves} \newline (24\%, n=385)
    \newline \quad \textit{- Expressed body image concerns} \newline \quad \textit{(14\%, n=229)}
    \newline \quad \textit{- Struggled with weight} \newline \quad \textit{(7\%, n=110)}
    \newline \quad \textit{- Showed self-loathing} \newline \quad \textit{(3\%, n=46)}
    & \textit{O: I really don't like my body and on top of that I come from a hairy male counterpart.} \newline \newline 
    \textit{P: I don’t look good in that at all. I have to lose weight!} \newline \newline 
    \textit{P: I just looked in the mirror and thought "how is being this ugly possible?"}
    & \multirow{6}{3.5cm}{\textbf{Giving Validation} (36\%, N=371) \newline \textbf{Shared Struggles} (24\%, N=263) \newline \textbf{Emotional Support} (17\%, N=189) \newline \textbf{Evading} (15\%, N=166) \newline \textbf{Encouragement} (8\%, N=88) \newline \textbf{No Response} (1\%, N=16)} \\ \cline{2-3}

    & \textbf{Attempted to modify their looks} \newline (18\%, n=295) 
    & \textit{P:I need teeth corrections, eye surgery, and a nose job before I'll be hot.}
    & \\ \cline{2-3}

    & \textbf{Experienced severe consequences of negative body image} \newline (10\%, n=157) 
    & \textit{P: O: When I first got my ED, I was eating 300-400 calories a day and exercising all the time.}
    & \\ \cline{2-3}

    & \textbf{Sought reassurance} \newline (6\%, n=100) 
    & \textit{P: What do you think of this pic? I’m not sure if it’s good enough to be posted.}
    & \\ \cline{2-3}

    & \textbf{Evaluated themselves against others} \newline (5\%, n=79) 
    & \textit{O: THE STUFF I’D DO TO LOOK LIKE THIS GIRL IS UNBELIEVABLE!}
    & \\ \cline{2-3}
    
    & \textbf{Shared personal experiences of body-shaming} \newline (5\%, n=78) 
    & \textit{P: Sometimes it’s just so hard being skinny. They always call me skinny girl at school and they say mean things like “your belly is so flat.”}
    & \\ \hline

    \multirow{3}{2.6cm}{\textbf{Teens body-shamed others} \newline (19\%, n=295)} 
    & \textbf{Judged and body-shamed others} \newline (14\%, n=217) 
    & \textit{P: I think you should gain more weight. Why are you Asians always so skinny?} \newline 
    \textit{O: Well, why are Latinos so fat in the first place?}
    & \multirow{3}{6cm}{\textbf{Criticizing Others} (71\%, N=210) \newline \textbf{Standing Up} (16\%, N=46) \newline \textbf{Evading} (10\%, N=31) \newline \textbf{No Response} (3\%, N=9)} \\ \cline{2-3}

    & \textbf{Body-shamed others to be funny} \newline (5\%, n=80) 
    & \textit{O: You’re some skinny shit} \newline \textit{P: It's ok I never really liked me either tbh.}
    & \\ \hline

    \multirow{3}{2.6cm}{\textbf{Teens promoted positive body image} \newline (13\%, n=207)} 
    & \textbf{Lifted others up} \newline (5\%, n=86) 
    & \textit{P: He was hot asf. He was tall, white, and tan with blonde hair and blue eyes!}
    & \multirow{3}{6cm}{\textbf{Appreciation} (48\%, N=100) \newline \textbf{Giving Validation} (18\%, N=37) \newline \textbf{Criticizing Others} (14\%, N=29) \newline \textbf{Evading} (11\%, N=23) \newline \textbf{No Response} (6\%, N=12) \newline \textbf{Standing Up} (3\%, N=6)} \\ \cline{2-3}

    & \textbf{Encouraged self-acceptance} \newline (5\%, n=77) 
    & \textit{O: You don’t need to be skinny to be pretty.}
    & \\ \cline{2-3}

    & \textbf{Critiqued societal standards} \newline (3\%, n=44) 
    & \textit{P: So that's why women change their appearance. They don't do it because they want to, but because they’re looked down on for being outside societal norms.}
    & \\ \hline

    \end{tabular}
    \label{RQ1&2}
\end{table}

 \subsection{Data Analysis Approach}
To address our first and second research questions, \edit{we utilized a mixed methods approach, combining qualitative thematic analysis with quantitative statistical analysis.} We used thematic analysis \cite{braun2012thematic} to explore key aspects of teens’ discussions about body image and/or body shaming,  as well as the responses they received to these disclosures. Initially, we familiarized ourselves with the data by coding conversations based on relevancy criteria, which allowed us to create initial codes and develop our primary research questions. The first author then open-coded a subset of sub-conversations to gain preliminary insights and establish the initial codebooks. These initial codes were reviewed with the last author and the larger research team to reach a consensus on coding the remaining sub-conversations. 
We also coded the conversations to determine whether they were one-to-one or group discussions and tracked the types of media files exchanged during these interactions. The first author then finalizes the codes with continuous guidance from the last author. After completing the coding process, the research team collaborated to organize the codes into the conceptual themes presented in this paper.

For RQ1, we identified three main themes in teens' discussions about body image with peers. For RQ2, we explored the type of responses they received for these discussions.
 Table \ref{RQ1&2} presents our final codebook for RQ1 and RQ2, featuring main themes, codes, illustrative quotations, and responses. 
 To answer our RQ3, we conducted a Chi-squared between-group analysis (X²) to examine any significant differences between codes based on whether the conversation was a dyad or group conversation. The X² test of independence is used to conduct between-group tests among two or more variables \cite{sharpe2015chi}. To demonstrate the significant differences among our codes, we used standardized residuals, calculated by dividing the difference between expected and observed values by the square root of the expected value \cite{sharpe2015chi}. Through this X² test, we aimed to interpret the nuanced differences in teens' body image disclosures and responses across dyadic and group conversations.

\section{Results}
In this section, we first summarize the descriptive characteristics of our sample, followed by a detailed discussion of the key themes from our qualitative analyses (RQ1 and RQ2). Finally, we present the statistical findings related to RQ3.
 In two-way conversations, `P' refers to the primary participant, and `O' denotes the other individual. For group conversations, 'O1', 'O2', 'O3', etc., were used to identify other participants. 

\subsection{Descriptive Characteristics of Our Sample}
Our final dataset consisted of 1,596 sub-conversations within 419 conversations from 67 unique participants (compared to the 195 total participants in the entire dataset). There was an average of 6 conversations (min = 1, max = 62), and 24 sub-conversations (min = 1, max = 315) per participant. Each conversation had an average of 4 sub-conversations per conversation (min = 1, max = 199). Of the 1,596 sub-conversations, 1,051 (66\%) were between participants and one other conversation partner, while 545 (34\%) were group conversations. Of the 67 teens, their ages ranged between 13 and 17 with a mean of 15.72 years and a median of 16 years. Most participants identified as female (64\%), followed by male (25\%), and those who did not specify their gender (10\%). The breakdown of participant race was as follows: 84\% (n=56) identified as White/Caucasian, 7\% (n=5) identified as Asian/Pacific Islander, 6\% (n=4) identified as Hispanic/Latino, 1\% (n=1) identified as Black/African-American, and 1\% (n=1) preferred to self-identify. In the following sections, we thoroughly present each disclosure theme (RQ1) in detail 
and examine the corresponding responses to these disclosures (RQ2) as shown in Table \ref{RQ1&2}.
\begin{table}[ht]
\centering
\caption{Type of Responses}
\begin{tabular}{|p{3.5cm}|p{3.5cm}|p{7cm}|}
\hline
\textbf{Themes} & \textbf{Codes} & \textbf{Illustrative Quotation} \\ \hline

\textbf{Peers offered support and showed appreciation} \newline (66\%, N=1,047) & 
\textbf{Giving validation} \newline (26\%, n=407) & 
\textit{P: I'm so ugly and worthless I don't deserve anything} \newline \textit{O:You deserve everything and you're none of that things you've said!} \\ \cline{2-3} 

& \textbf{Sharing similar struggles} \newline (16\%, n=263) & 
\textit{P: I hope I still get to wear masks even after the pandemic, that shit covers my ugly face} \newline \textit{O: me too! masks have been a lifesaver} \\ \cline{2-3} 

& \textbf{Emotional support}\newline (12\%, n=189) & 
\textit{P: my parents are starting keto and them talking about calories hurt me so much} \newline \textit{O: Aghh Dude this is pain. That sucks, I'm sorry} \\ \cline{2-3} 

& \textbf{Appreciation} \newline (6\%, n=100) & 
\textit{P: I know someone cute when I see that person and I think you’re cute} \newline \textit{O: You’re too sweet. And you’re really gonna make me blush here. Thank you} \\ \cline{2-3} 

& \textbf{Encouragement}\newline (6\%, n=88) & 
\textit{O1: I don’t know how to be on my own. but i’m doing workouts daily now} \newline \textit{O2: That’s good! It’ll carry on. And it’ll show!} \\ \hline

\textbf{Peers respond with criticism rather than defense} \newline (18\%, N=293) & 
\textbf{Criticizing others} \newline (15\%, n=239) & 
\textit{P: I just realized that she looked ugly asf in that picture} \newline \textit{O: Yeah, the straight hair is ugly} \\ \cline{2-3} 

& \textbf{Defending others} \newline(3\%, n=54) & 
\textit{O1: y r u wearing those pants. I feel they’re not appropriate for your weight} \newline \textit{O2: You need to mind your business like who tf asked you how you felt about her pants?} \\ \hline

\textbf{Peers overlooked or shifted the topic} \newline (16\%, N=256) & 
\textbf{Evading} \newline (14\%, n=219) & 
\textit{P: No idea how guys stay skinny and eat so much shit. It don’t make sense. How does that work??} \newline \textit{O: BRUH THAT SONG} \\ \cline{2-3} 

& \textbf{No response} \newline (2\%, n=37) & 
\textit{P: I’m not pretty and I know. Plus you would get annoyed with me and bored too easily} \newline \textit{P: Stop leaving me on read meanie!} \\ \hline
\end{tabular}
\label{RQ2Themes}
\end{table}

 \subsection{Teens Disclosed Their Struggles With Negative Body Image: (RQ1)}
When teens and their conversation partners discuss body image, they most often disclosed their struggles with negative body image (68\%, \textit{n = 1,095}). In the following sections. we go deeper into the different types of their struggles. 
\subsubsection{Teens Judged Themselves.}
A substantial number of instances (24\%, \textit{n = 385}) involved teens \textbf{judging themselves} based on their own appearance, particularly regarding their looks and weight. Among these conversations, a common behavior we observed was that teens frequently \textit{expressed body image concerns} (14\%, \textit{n = 229}). In these conversations, teens often shared their concerns and insecurities about their appearance, focusing on various aspects of their bodies and how they wished to look. \edit{We also noticed that once these concerns emerged, they often resurfaced over time in later interactions during conversations with different people}. We saw that these anxieties intensified after others took screenshots of them, particularly around perceived flaws like facial features, skin discolorations, and spots. In many cases, teens also shared pictures of themselves that highlighted these perceived imperfections, 
expressing disappointment in their current appearance by comparing it to how they used to look in the past, such as the message below.
\begin{quote}
    \textit{``P: Hopefully I'll soon look more attractive once all my acne clears up. I'd look so much better. It's such a small thing, but it really messes with how I feel about myself every time I look in the mirror"} - Female, 17 years old
\end{quote}


Some other conversations (7\%, \textit{n = 110}) depicted how \textit{teens struggled with their body weight}. 
Teens primarily self-criticized their weight and how it affected their self-esteem and daily life, such as blaming themselves for gaining weight, not sticking to a planned diet, or failing to exercise regularly. Teens often expressed dissatisfaction with their bodies, often referring to themselves as "fat" (e.g. \textit{``: Looked in the mirror, and now I wanna cry because I’m built like and as fat as a pumpkin"} - Female, 15-year-old). In a few instances, teens also judged themselves for being too skinny (e.g. \textit{``P: I hate that my legs look skinny in these shoes"} - Female, 16-year-old). In both instances of weight-related self-judgment, whether teens perceived themselves as too heavy or thin, the inability to wear desired clothing often emerged as a trigger. The frustration of not fitting into or feeling comfortable in these outfits seemed to strengthen their negative feelings about their weight, leading to a cycle of body dissatisfaction and self-criticism.
\begin{quote}
    \textit{``P: Sadly there is no way I could just wear the shirt because it hugs my body and I don’t like my stomach"} - Female, 15-years-old
\end{quote}

The last behavior we observed when teens judged themselves was a tendency to engage in extreme \textit{self-loathing} (3\%, \textit{n = 46}). In these cases, teens frequently used disparaging language, often referring to themselves as "ugly." This extreme form of self-criticism was often rooted in deep feelings of inadequacy and low self-esteem, which many teens attributed to their appearance. 
Teens' negative self-perception was not just a reflection of their appearance but also intertwined with their struggles in forming relationships, which they saw as proof of their unattractiveness. 
\edit{
\begin{quote}
    \textit{``O: I wanna date people but i'm UGLY AS BRUSSUL SPROOTS”} -Conversation with 13-years-old, female
\end{quote} 
}
\subsubsection{Teens Attempted to Modify Their Looks.}
Besides judging themselves, we found that in around one-fifth of the conversations, we observed that teens \textbf{attempted to modify their looks}, whether self-imposed or influenced by others  (18\%, \textit{n = 295}). A significant portion of these discussions focused on their efforts to \textit{lose weight} (13\%, \textit{n = 209}). Teens shared a range of strategies, including exercising, following specific diet plans, and tracking their calorie intake. This focus was reflected in the media they shared, such as photos of fitness models they admired, clothing they hoped to wear, and personal snapshots documenting their progress and efforts toward achieving their goals.
These discussions also revealed diverse motivations for weight loss. Some conversations by teens \edit{\textit{(n = 37})} were driven by concerns unrelated to their body image such as potential health issues, like becoming prediabetic, while others \edit{\textit{(n = 20})} were motivated by a desire to appear more attractive. In certain instances \edit{\textit{(n = 10})}, teens expressed that losing weight was necessary to feel confident enough to date, believing that being slimmer was crucial. For others \edit{\textit{(n = 8})}, the motivation came from wanting to fit into specific outfits or trendy clothing only available in smaller sizes. The pressure to conform to certain body types, especially in the context of fashion, played a significant role in driving these weight loss efforts.
\begin{quote}
\textit{
``P: I should buy more dresses. Sometimes I feel like losing weight so I could fit in clothes that are super pretty cause sometimes they’re like only small sizes"} -Female, 17-years-old
\end{quote}

Another way teens and their conversation partners attempted to modify their looks was by concealing perceived problem areas (5\%, \textit{n = 86}). Teens frequently expressed anxiety and insecurity about how their appearance would be judged by others, both online and offline, which often led them to adopt behaviors aimed at hiding or minimizing specific areas of concern. 
Various approaches were discussed to address these insecurities. Some discussions by teens \edit{\textit{(n = 9})} talked about how they turned to makeup as a means of masking perceived imperfections.
For others \edit{\textit{(n = 11})}, the possibility of cosmetic surgery was raised as a more permanent solution to their concerns, with discussions revolving around procedures to alter facial features, enhance body shape, or eliminate perceived flaws entirely. 
Teens often relied on social media filters to alter their appearance in photos, creating an idealized version of themselves that aligned more closely with societal beauty ideals. 
In addition, teens shared in some conversations \edit{\textit{(n = 5})} how they used clothing to conceal areas of their bodies they were self-conscious about. Loose-fitting clothes, baggy shirts, or specific styles were worn to hide what they considered problematic areas, such as their body shape. For some, the clothing became a protective barrier, allowing them to feel less exposed to potential judgment.

\begin{quote} \textit{``P: I wear baggy shirts because I'm insecure about how I look"} -Female, 15-years-old
\end{quote}

\subsubsection{Teens Shared Severe Consequences They Faced for Negative Body Image.}

In other instances, there were some conversations where we found teens and their conversation partners shared about the \textbf{severe consequences they faced for negative body image} (10\%, \textit{n = 157}). In these cases, we found that teens mostly (8\%, \textit{n = 126}) engaged in discussions related to their \textit{eating disorder}. Throughout these discussions, teens shared extreme dieting practices to lose weight. \edit{For instance, as conversations evolved, we observed a recurring theme where some discussions were specifically dedicated to encouraging extreme eating behaviors, such as extended fasting, sometimes lasting up to 24 hours.} These conversations often took place within groups that actively supported such behaviors \edit{\textit{(n = 33})}.
They also discussed their daily calorie intake \edit{\textit{(n = 22})}, frequently mentioning shockingly low numbers, such as staying under 500 calories a day. Teens admitted that these behaviors, beyond facilitating weight loss, boosted their mood and provided a sense of control. Despite reaching their weight goals, many did not plan to stop these behaviors. Instead, they often sought out sources of inspiration to continue their eating disorder journey, such as constantly checking "thinspo" images for motivation, which further discouraged eating and promoted food restriction. 
In addition to these behaviors, teens shared images of substitutes for food, such as energy drinks consumed instead of meals. Memes were commonly shared as well, humorously depicting eating disorders such as anorexic skeletons at a buffet or jokes about breaking a fast. Teens also shared screenshots of conversations where they challenged each other to see who could survive without food for the longest period, further reinforcing these harmful behaviors. Additionally, we observed that teens faced a dilemma when sharing updates related to eating habits associated with eating disorders. While they sought attention and support, they also feared that their disclosures might raise a concern or prompt intervention (e.g. \textit{P: I want to tell my friends about my ED but I'm worried they will make me stop or worry about me"} - 16-year-old, Female).
Teens also expressed frustration when their families showed concern, as this could lead to forced recovery. Teens admitted that their families would monitor their eating habits and sometimes hide scales to prevent obsessive weight monitoring. To avoid detection, we observed that teens adopted strategies to hide their eating disorders, such as pretending to eat, using laxatives, or inducing vomiting after meals.
\begin{quote} { \textit{ ``O: I am good at eating an entire pizza and then vomiting it up to stay skinny"} -Conversation with 13-years-old, female 
} \end{quote}

Another consequence of negative body image we observed was when teens and their conversation partners discussed \textit{mental health problems} (2\%, \textit{n = 31}). In these conversations, teens shared how their negative body image seemed to affect their mental well-being, contributing to feelings of anxiety, low mood, and, in some cases, suicidal thoughts. Many expressed ongoing emotional struggles tied to how they viewed themselves, noting that their self-perception had a direct impact on their overall mood and mental state. These negative feelings often extended beyond body dissatisfaction, influencing their approach to social interactions and relationships. 
\subsubsection{Teen Sought Reassurance}
Apart from that, we found some instances where teens \textbf{sought reassurance} from others. In approximately 6\% (\textit{n} = 100) of the conversations, teens exhibited significant concern regarding how others perceived them, frequently seeking reassurance to strengthen their confidence about their appearance. For example, some teens directly sent personal photos to friends or peers requesting feedback on their looks \edit{\textit{(n = 9})}. Further, this need for affirmation extended to their social media posts, where the accumulation of likes and comments served as a form of social currency. When posts failed to attract the anticipated level of engagement, many teens actively reached out to friends, requesting them to like or comment on their posts.
\begin{quote}
    {
    \textit{
  ``P: Will you comment something on my recent post?  I don't like it when people don't comment I'm sorry lol"} -Female, 16-years-old
    }
\end{quote}

\subsubsection{Teens Evaluated Themselves Against Others.}
We also found several instances where teens \textbf{evaluated themselves against others} (5\%, \textit{n = 79}). These conversations showed a tendency among teens to compare their appearance with that of others, which often appeared to contribute to dissatisfaction with their own looks. This comparison occurred not only with individuals they were directly interacting with, such as peers and family members, but also extended to public figures and posts encountered on social media. Teens shared images of people with body types or outfits they admired, indicating a desire to look similar. These images seemed to serve as points of reference for how teens assessed their own appearance, leading to feelings of inadequacy when they felt they did not meet those standards. 
\begin{quote} { \textit{``O: THE STUFF I’D DO TO LOOK LIKE THIS GIRL IS
UNBELIEVABLE!"}-Conversation with 15-years-old, male } \end{quote}
\subsubsection{Teens Shared Personal Experiences with Body-shaming}
Lastly, in approximately 5\% (\textit{n = 78}) of the conversations, teens \textbf{shared their body-shaming experiences}. These experiences were diverse and they happened both online and offline, with some teens being shamed for their weight, while others faced comments targeting their physical appearance, race, or skin color. In those occasions, teens expressed feeling uncomfortable with their bodies as a result of such remarks, which appeared to affect their self-esteem. 
We also saw that a notable portion of these conversations \edit{\textit{(n = 13})} involved teens discussing body-shaming comments from their own families, particularly parents. Teens shared that their parents would often make remarks about their weight or highlight perceived flaws, which added a layer of complexity to their body image struggles. These family-related comments seemed to have a lasting impact, as the teens navigated conflicting emotions of familial care and criticism.

\begin{quote} { \textit{``P: I have one top, and my mom says she's not body-shaming me, but she said that my body is too big for the shirt, and I beg to differ!"} -Female, 16-years-old
} \end{quote}



\subsection{Peers Mostly Offered Validation to Teens' Struggles With Negative Body Image (RQ2)}

When teens and their conversation partners disclosed their struggle with negative body image, the responses they received from peers were mostly supportive. As illustrated in Table~\ref{RQ1&2}, we observed that peers mostly \textbf{gave validation} to these types of disclosures. 
  Teens mostly received such validations from their peers when they exhibited self-judgment or sought reassurance. As such, these responses often involved affirming the individual’s appearance, countering negative self-perceptions, and addressing insecurities, ultimately offering reassurance and boosting confidence. In most of these conversations, we noticed that teens frequently shared about their negative perceptions of their own body images and received light-hearted responses from their peers that validated them by downplaying their insecurities.
\begin{quote}
    {
    \textit{
  ``P: I'm ugly as fuck"} -Female, 15-years-old \newline
\textit{
``O: No you're not. At worst you're average! come to my school and you’ll feel like a supermodel"}
}
\end{quote}
\edit{However, we observed that this type of response did not seem to stop teens from sharing their negative body perceptions over time, as they repeatedly revisited the same concerns in both the same and subsequent conversations.}


Another type of responses that was also common when teens disclosed their struggles with negative body image was centered on \textbf{sharing similar struggles}.
In these instances
, peers responded by recounting their own personal challenges. This was especially common when teens discussed their attempts to modify their looks, mainly when trying to lose weight. Through these shared stories, teens appeared to find comfort in realizing they were not alone in their difficulties, which seemed to foster a sense of solidarity and mutual understanding. Many of these interactions involved detailed personal stories, where peers offered relatable examples from their own lives. For instance, when one teen expressed frustration about regaining weight, another responded with their own similar experience, creating a moment of connection:
\begin{quote} 
\textit{
``P: I’ve gained all the weight back from summer"} Female, 13-years-old\newline 
\textit{``O: Literally me, I need someone to hold me accountable"} \end{quote}


We also saw that some of the responses involved peers offering \textbf{emotional support} as a response.
In these dialogues, teens provided comfort and understanding, especially to those who expressed self-judgment or experienced body-shaming. These interactions typically included empathetic remarks, acknowledging the distress shared by their peers. Rather than offering solutions, teens focused on making their peers feel heard and supported through expressions of sympathy. These exchanges included both brief acknowledgments and more thoughtful reflections, which seemed to help peers feel that their emotions were being validated. 
\begin{quote}
    \textit{
    ``P: welcome back to, I feel fat! the daily segment of my life, normally happening between the hours of 6-10pm, where I look in the mirror and physically gag"} Female, 16-years-old\newline
    \textit{``O1: Aw. Sorry you feel like that"\newline
    ``O2: Ohh I'm a big fan of this one"
    }
\end{quote}


 In other responses to teens' disclosures about their negative body image, we observed that some of these interactions lacked meaningful engagement. For instance, we saw that sometimes
 peers appeared to be \textbf{evading} the subject and steering the conversation away from body image toward other topics 
 . The reasons for this evasion were not always clear, but it was most common when self-judgment was present in the conversation. In these responses, conversation partners often felt uncomfortable with the topic or reluctant to engage in sensitive discussions surrounding negative body image. \edit{Over time, as self-judgment persisted among teens, conversation partners started pulling away, not necessarily out of indifference but from the exhaustion of repeatedly offering reassurance. When the same individual continuously expressed self-doubt, it became increasingly difficult for peers to stay engaged, making support feel less sustainable and more draining.}
As a result, instead of addressing the body image concerns directly, they often redirected the discussion to more neutral or unrelated subjects, leaving the teen without the support they were probably looking for.
\begin{quote}
    \textit{
    ``O: if my legs were slender and skinny, they would look long and pretty" 
"}\newline
    \textit{``P: I'm going out tomorrow. It's the first time I've gone out in like 3 weeks"
    } Female, 15-years-old
\end{quote}

In other instances of the responses 
, teens and their peers exchanged \textbf{encouragement}. In these interactions, peers mostly motivated and supported teens in their journeys to modify their looks by losing weight or when experiencing severe consequences of negative body image by sharing about their eating disorders. Throughout these responses, we saw that they often provided words of encouragement, praising the efforts and progress made by the others. On one hand, these responses were encouraging unhealthy behaviors like eating orders. On the other hand, peers also exchanged advice on overcoming challenges such as maintaining healthy habits and managing diet and exercise routines when trying to lose weight. For instance, peers offered tips on creating sustainable meal plans, shared effective workout routines, and suggested strategies to stay motivated. Peers also emphasized the importance of consistency and patience, reassuring teens that progress might be slow but that their efforts would eventually pay off. 
\begin{quote}
\textit{
``P: Have you been active recently?"} -Female, 17-years-old\newline
\textit{``O: No, I don’t know how to be on my own. But I am doing workouts daily now."\newline
``P: That’s good! It’ll carry on and it’ll show!"
}
\end{quote}

Finally, we observed that in a few 
responses to these disclosures
, peers provided \textbf{no response} at all. There are endless possibilities for why this occurred. However, one theme we noticed was that peers became less responsive over time, not necessarily due to exhaustion, but possibly because they felt their support was ineffective. When teens repeatedly sought reassurance, peers may have started to doubt whether their support or validation was making a difference, leading to uncertainty about how to continue the conversation. This uncertainty could have resulted in silence, as peers might have felt stuck between wanting to help but not knowing how to offer meaningful support.
\begin{quote}
    \textit{
    ``P: That's if you can put up with my ugly ass face.} Female, 15-years-old \newline
    \textit{P: I'm sorry I'm probably bothering you a lot.\newline
    P: Sorry, I will leave you alone. I hope you had a great day and have a great night!"
    }
\end{quote}

\subsection{Teens Body-shamed Others (RQ1)}
While the majority of conversations showed teens disclosing their struggles with negative body image, around one-fifth of the conversations (19\%, \textit{n = 295}) had instances where teens body-shamed others. In the sections that follow, we explore this topic further.
\subsubsection{Teens Judged and Body-shamed Others.}
Teens and their conversation partners often (14\%, \textit{n = 217}) exchanged \textbf{body judgment and shaming} comments about others'  appearances. In these conversations, teens often engaged in gossiping about others' social media posts, where teens took and shared screenshots of others and harshly critiqued physical features. \edit{We also observed how these discussions evolved, gradually becoming a way for teens to bond, with body shaming emerging as a common thread that connected them. As conversations progressed, criticism of others often shifted into a means of reinforcing group cohesion, allowing those present to feel validated by positioning themselves as superior in comparison.}
Teens and their conversation partners seemed particularly comfortable expressing these criticisms about people they did not personally know, especially in online environments where the perceived distance reduced the immediate consequences of their remarks. This encouraged them to make comments they might otherwise avoid in face-to-face interactions. Moreover, we saw that teens occasionally justified their critical remarks by framing them as expressions of honesty, using this as a rationale for comments that could be hurtful \edit{\textit{(n = 14)}}. This tendency further normalized such behaviors, reinforcing harmful norms around appearance and criticism in their interactions.
\begin{quote} 
\textit{
``P: Did you see her latest post? She really shouldn’t be wearing that dress”} -Male, 14-years-old  \newline
\textit{``O1: I know, right? It doesn’t suit her at all"  \newline
``O2: I’m just being honest. Someone needs to tell her she doesn’t have the body for that outfit!"}
\end{quote}
\subsubsection{Teens Used Body-shaming Humorously.}
We found instances where teens and their conversation partners used \textbf{body shaming for the sake of being funny} (5\%, \textit{n = 80}). Although intended to be playful, these remarks still involved negative comments about physical appearance and, therefore, could be classified as body shaming. Teens often made light-hearted remarks about features such as weight, height, or other physical traits, presenting their comments in a humorous context intending to elicit laughter and create a shared sense of fun. However, we saw that these jokes often blurred the line between harmless banter and actual body-shaming, sometimes leading to confusion or discomfort among those involved. While these remarks may have been framed as humorous, their underlying impact could be more serious. Teens involved in these exchanges may not always have been aware of the emotional toll their comments had on others, especially when the humor targeted sensitive aspects of physical appearance. 

\begin{quote} \textit{``O1: You’re so short, we’re gonna have to put you on a step stool for the group photo! Just messing with you\newline ``P: Yeah, or we’ll just him you around in our backpacks"} -Male, 15-years-old  \newline \textit{``O2: LOL Seriously, how do you even reach the top shelf at home?
"}
\end{quote}

\subsection{Peers Mostly Responded With Criticism Rather than Defense to Body-shaming (RQ2)}
When teens and their conversation partners body-shamed others, the reactions they received from peers varied broadly. However, our analysis revealed that the majority 
of responses had negative tones 
with teens \textbf{criticizing others} as shown in Table~\ref{RQ1&2}.
We saw that these interactions frequently occurred in conversations where body shaming was present, sometimes even when it was introduced humorously. We saw that peers responded to body judgments or shaming by mostly joining the criticism rather than defending the targeted individuals. These conversations often escalated as other group members joined in, particularly when body shaming was framed as humor. In these cases, teens tended to support the behavior by laughing or adding to the ridicule, especially when the discussion focused on individuals not present in the conversation.

\begin{quote}
    \textit{
    ``O1: She's so weirdly shaped. look at her
"}\newline
    \textit{``O2: Yess. she has angry eyebrow  lol"
    }\newline
    \textit{``O3:`OMG. you're right!"
    } -Conversation with 16-years-old, male
\end{quote}


In a smaller portion of the responses to the body shaming disclosures
, peers \textbf{stood up} to defend others. These instances primarily took place in group settings where body shaming was part of the conversation where we saw that peers actively stepped in to stop the judgment, whether aimed at individuals within the conversation or others not present. Peers often challenged negative comments, redirecting the conversation toward a more respectful tone. For example, when someone made an offensive remark about another person's appearance, others would quickly intervene, reminding the group of the need for kindness and reinforcing that everyone deserved respect and validation regardless of their looks.

\begin{quote}
    \textit{
    ``O1: You CALL YOURSELF UGLY to FISH for compliments. But you’re still FAT! Anyway, go eat"}\newline
    \textit{``O2: Bro. Don't say that about him. Fat shaming isn’t funny!"
    }
\end{quote}

\subsection{Teens Promoted Positive Body Image (RQ1)}
Lastly, we saw that some of the conversations showed teens promoting positive body image (13\%, \textit{n = 207}). In the following sections, we dive deeper into that.

\subsubsection{Teens Lifted Others Up.}
Within these positive conversations, we saw that 5\% (\textit{n = 86}) was about teens \textbf{lifting up others}.  In these cases, teens mostly complimented the appearance of others, including both the person they were chatting with and those external to the conversation. we saw that teens frequently praised the appearance of their friends and peers. They highlighted specific features they admired, such as a friend's smile, hair, or fashion sense, and expressed genuine appreciation for their appearance. We saw that this practice of complimenting each other not only helped to promote a positive atmosphere but also strengthened their bonds. These holistic compliments, which sometimes acknowledge both inner qualities and outer appearances, exemplify how teens uplift each other in their private messages. For instance, one 17-year-old female expressed:
\begin{quote}
    {
    \textit{
``P: I like, your personality, the way you look and your sense of humor.   BEAUTIFUL!"} -Female, 17-years-old
}
\end{quote}
\subsubsection{Teens Encouraged Self-acceptance.}
We also observed instances where teens \textbf{encouraged self-acceptance} in various ways (5\%, \textit{n = 77}). In these discussions, teens often emphasized the importance of feeling confident and making choices that enhance personal well-being, whether through embracing their natural appearance or using cosmetic enhancements if it helped boost self-esteem. 
Teens also shared practical strategies for fostering a positive body image. These included shifting focus from weight loss to developing healthy habits, such as maintaining a balanced diet, exercising for strength and well-being, and engaging in activities that bring joy and fulfillment. 
Additionally, in a few cases \edit{\textit{(n = 17)}}, teens encouraged participation in positive social media challenges aimed at spreading body positivity, such as tagging friends they found beautiful or sharing unfiltered photos of themselves. These challenges fostered a sense of community and support, helping others build confidence by receiving and offering affirmations in a more inclusive and uplifting environment.
\begin{quote}
    \textit{
    ``P: Some women find it easy to criticize each other. With all the negativity going around, let’s do something positive!! Upload one picture of yourself... ONLY YOU. Then tag 10 more beautiful ladies to do the same. COPY and PASTE. If I tag you... don’t disappoint me. Let’s spread some positivity"} -15-years-old, Female
\end{quote}
\edit{Lastly, we observed that these positive discussions remained a central focus of this type of conversation over time, indicating that the entire group chat seemed to serve as a dedicated space for body positivity. The recurring themes of self-acceptance and confidence suggest a strong community built around empowerment and inclusion.}
\subsubsection{Teens Criticized Societal Standard.}
Lastly, in some conversations, teens expressed \textbf{critiques of societal standards} (3\%, \textit{n = 44}). Many of these discussions centered on their perceptions and criticisms of societal beauty standards, which often prioritize certain physical features and races. Teens frequently voiced frustration with these narrow ideals, emphasizing the importance of recognizing and celebrating diversity. They argued against the notion that videos or content are necessary to showcase the beauty of different races, asserting that all races have unique physical features that make them special. Another key point of discussion within these conversations was the concept of being "healthy skinny" versus "risky skinny," with teens debating what constitutes a healthy body image and how societal pressures can promote unhealthy standards of thinness. Additionally, they observed that beauty standards are not static but evolve over time, noting that what is considered attractive today might have been viewed negatively in the past. 

\begin{quote} \textit{``P: It's just so stupid; why should everyone have to fit into the same beauty standard when diversity and different features are what make everyone attractive? It never makes sense" } -Female, 16-year-old\end{quote}

\subsection{Peers Mostly Showed Appreciation to Positive Body Image Promotion (RQ2)}
When teens promoted positive body image, the responses they received were predominantly positive. As highlighted in Table~\ref{RQ1&2}, \textbf{appreciation} emerged as one of the most common types of responses
. 
In these instances, we saw that teens and their peers exchanged words of appreciation more often in response to the conversations where others lifted them up by offering compliments or when others encouraged them to accept themselves. We saw that teens frequently expressed gratitude towards their peers for making them feel better. This appreciation also followed after peers shared motivational content that helped them navigate their body image challenges.
\begin{quote}
\textit{
``P: Your comparison weight loss video popped up when I was feeling the lowest I’ve felt in a while. Thank you so much for the motivation"} Male, 15-years-old\newline
\textit{``O: You’re welcome"
}
\end{quote}

We also observed that in some instances 
that peers responded to these positive body image remarks by \textbf{giving validation} in return
. In these cases, rather than merely acknowledging the initial comments, peers actively reciprocated by offering compliments in response. This exchange was particularly noticeable when the original disclosures centered around uplifting or encouraging others. By responding with compliments, peers reinforced the positive sentiments being shared, creating a pattern of mutual validation. 

In other cases, we saw that \textbf{criticism}
sometimes emerged as a response
, particularly to remarks challenging social beauty standards. In these exchanges, the conversation occasionally took on a more negative tone, as criticisms of these standards prompted some peers to shift the focus toward blaming specific individuals or groups seen as responsible for perpetuating such ideals.

\begin{figure}[htbp]
  \centering
\includegraphics[width=0.5\textwidth]{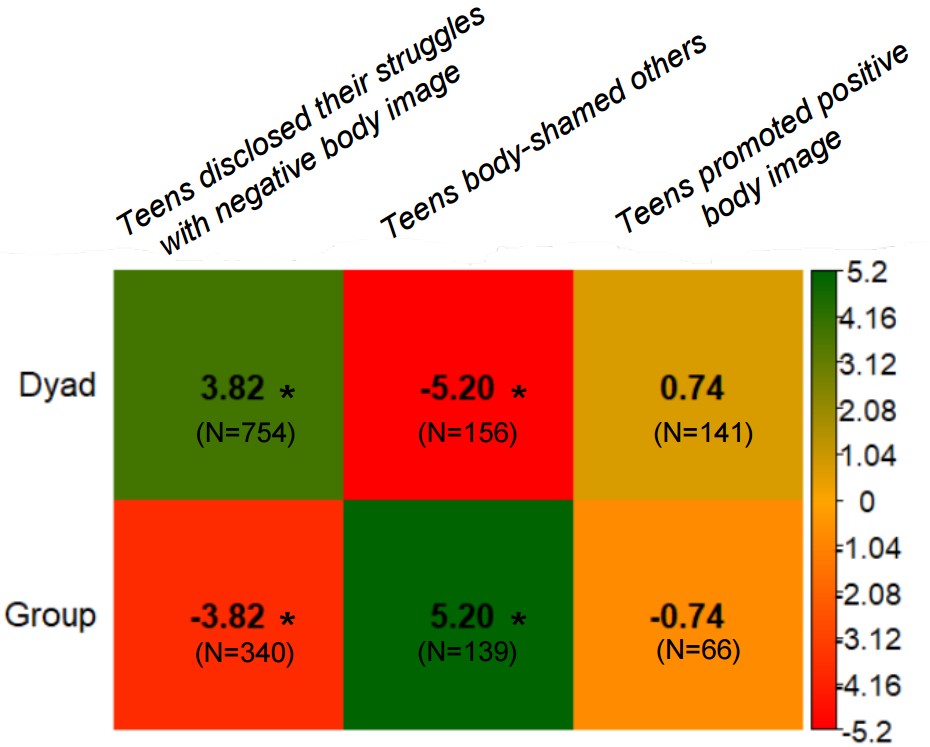}
   \caption{Results (standardized residuals) of the between-group analysis of the between-group analysis of teens' body image disclosures, based on whether the conversation occurred in a dyad or group setting. (*) indicates significant association. Note that green denotes a positive association, while red denotes a negative one.}
   \label{Chi_RQ1}
 \end{figure}

\subsection{Differences in Body Image Discussions: One-on-one vs. Group Conversations (RQ3)}
Body image disclosures and responses between teens and their peers varied significantly depending on whether the
conversations occurred in a one-on-one (dyad) or group setting. To explore these differences, we employed a chi-square ($\chi^2$) test, which revealed statistically significant differences. A detailed analysis of these findings is provided below.

\subsubsection{Body Image Disclosures Across One-on-one vs. Group Conversations}
Body image disclosures and responses between teens and their peers varied significantly depending on whether the conversations occurred in a one-on-one (dyad) or group setting. To explore these differences, we employed a chi-square ($\chi^2$) test, which revealed statistically significant differences. 
As shown in Figure \ref{Chi_RQ1}, the $\chi^2$ test indicated significant differences in the distribution of body image disclosures between the different body-image disclosures, based on relative expected proportions relative to the total number of conversations $(\chi^{2} (df=2)=27.12, p < 0.001$). 
We found that the relative proportion of the conversations where teens shared their struggles with negative body image 
reached positive significance in dyad settings, indicating teens were significantly more likely to disclose their struggles with negative body image in private dyadic conversations compared to the group chats. In contrast, such disclosures were less likely to take place in group settings, showing a significant negative association. 
This suggests that one-on-one conversations seemed to offer teens a confidential, supportive environment, where they felt more comfortable sharing deeply personal concerns. 
\begin{quote}
    \textit{
    ``P: I'm horrible and ugly. I'm worthless I don't deserve anything
    "} Female, 17-year-old \newline
   \textit{`` O:  You do deserve everything and you're none of that things you've said. You're pretty"}
\end{quote}

Aside from teens disclosing their struggles, we observed that teens were significantly less likely to engage in body shaming in dyadic conversations, revealing a clear negative association, as illustrated in Figure \ref{Chi_RQ1}. 
However, in group settings, we saw that teens were more likely to engage in body shaming, showing a strong positive association. The group context likely fostered these behaviors as we observed teens, under social pressure, seeking to align with group norms. Additionally, we saw that once body-shaming started, it had a tendency to escalate, with many teens joining in, which kept the behavior going and made it last longer in these settings. 
\begin{quote}
    \textit{
    ``O1: Did you see how big her arms looked in that picture
    "}  \newline
    \textit{``P: Yup, and don't get me started on her outfit, definitely wasn't helping lol"} -Female, 16-year-old \newline
   \textit{``O2:  Honestly, she’s always wearing stuff that doesn’t suit her body"}
\end{quote}

\begin{figure}[htbp]
  \centering
\includegraphics[width=0.55\textwidth]{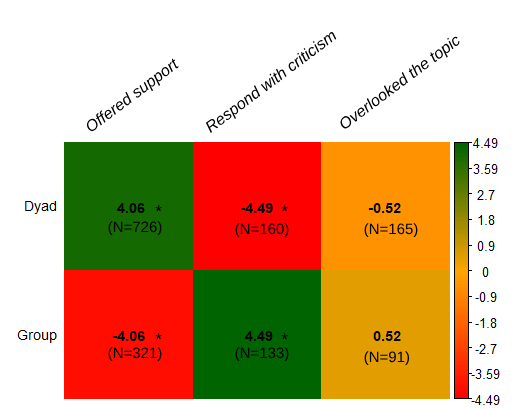}
   \caption{Results (standardized residuals) of the between-group analysis of the between-group analysis of teens' responses to body image disclosures, based on whether the conversation occurred in a dyad or group setting. (*) indicates significant association. Note that green denotes a positive association, while red denotes a negative one.}
   \label{Chi_RQ2}
 \end{figure}

\subsubsection{Body Image Responses Across One-on-one vs. Group Conversations}
As illustrated in Figure \ref{Chi_RQ2}, the $\chi^2$ test indicated significant differences between the responses based on the expected proportions relative to the total number of conversations $(\chi^{2} (df = 2) = 22.36, p < 0.001)$. 
In dyad settings, teens were significantly more likely to receive support when disclosing their body image, showing a strong positive association. In contrast, group settings were negatively associated with offering supportive responses as shown in Fig \ref{Chi_RQ2}. Teens in these private settings were frequently reassured, indicating that dyads may provided a comfortable space for sharing vulnerabilities and receiving personal, direct feedback where the attention was focused solely on them, without the distractions or pressures of a group setting. 
\begin{quote}
   \textit{
   O: even though you call yourself ugly you are so beautiful inside and out. I hope you find someone who will make you see the pretty girl that you are. Stay safe 
    } -Conversation with 15-years-old, female
\end{quote}
In contrast, group settings were negatively associated with offering supportive responses. 
Teens in groups were significantly less likely to receive direct support, possibly because the presence of multiple participants diluted the personal attention often needed when discussing sensitive topics like body image. The group interactions in our dataset appeared to result in less focused, personalized responses, with supportive responses being less common. This suggests that group settings may not provide the same level of individualized support that teens are more likely to experience in one-on-one conversations.

Another type of response that showed significant positive or negative associations was when teens engaged in argumentative exchanges, often by criticizing others. 
Teens were more likely to engage in such reactions in group conversations, displaying a significant positive association as demonstrated in Figure \ref{Chi_RQ2}. Teens in these group settings were more inclined to gossip about the appearance of others who normally were not part of the conversation. Also, managing multiple voices and perspectives in a group may have made it difficult to maintain a consistently supportive tone, leading to interruptions, disagreements, and, in many cases, a more judgmental atmosphere, increasing the likelihood of negative interactions.
\begin{quote}
   \textit{
 O1: You're all fat! \newline
O2: Seriously? At least I’m actually trying to lose weight, unlike you! \newline 
O3: Shut up O1, you're not only fat, but you're also the most insecure person here, always dragging others down to feel better about yourself!
    }-Conversation with 16-years-old, female
\end{quote}
In contrast, dyadic conversations were less likely to involve disagreement or criticism, showing a significant negative association. The personal nature of these one-on-one settings appeared to create an environment where teens received more focused, constructive, and supportive feedback. Without the distractions of multiple participants, dyadic conversations were generally more engaged and responsive, with fewer instances of disengagement or unsupportive comments. Teens likely felt more comfortable in dyads, knowing they were speaking directly to one person, which reduced the likelihood of facing harsh criticism.

\section{DISCUSSION}
In this section, we describe the implications of our findings in relation to prior work and provide
design implications.
\subsection{The Double-Edged Sword of Private Social Media Exchanges: Balancing Support and Risk (RQ1)}
\label{5.1}
Our study revealed that teens' private disclosures were heavily skewed toward sharing struggles with negative body image.
 Previous research consistently identified social media platforms as significant contributors to body image concerns among adolescents \cite{vandenbosch2022social, saiphoo2019meta, faelens2021relationship}. Based on this, we initially expected to find clearer evidence of social media's influence within these private discussions. However, the specific role of social media in shaping adolescents' body image struggles remained less apparent in the teens' conversations.
 \edit{Some of our findings did indicate that teens evaluated themselves against others, a tendency that might have been shaped by the algorithmically curated feeds they encountered on the platform. While teens tend to perceive themselves as in control of their feeds through curation and deliberate engagement with preferred content~\cite{wang2022tiktok,mcdonald2024me}, they may not fully grasp how much the algorithm already knows about them or the extent of data being collected~\cite{goray2022youths}. When unexpected content appears in their feeds, teens have been documented to ignore it rather than question why it was recommended to them~\cite{wang2022tiktok, mcdonald2024me}. As a result, they may overestimate their influence on personalization and underestimate the broader implications of algorithmic curation. As teens continue interacting with their feeds, algorithms may subtly reinforce negative body image without their direct awareness, exposing them to content they may struggle to control or avoid~\cite{milton2023see}. This concern is further amplified by the limitations of current social media feed algorithms, which often fail to capture nuanced user preferences~\cite{feng2024mapping}. Even when algorithms successfully personalize content, concerns persist regarding their potential to constrain teens' exposure to diverse and body-positive representations, which they may not proactively seek~\cite{mcdonald2024me}.
We believe this observation is not only exclusive to Instagram but can be extended to other image-based social media platforms, which are equally susceptible to fostering negative body image discussions due to their inherently visual nature~\cite{manikonda2017modeling}. For instance, Snapchat, with its widely used augmented reality (AR) filters, has been shown to contribute to body dissatisfaction by encouraging users to modify their facial and physical features, sometimes to unrealistic standards, raising concerns about the perpetuation of unattainable beauty ideals~\cite{bonner2023filters}. Similarly, TikTok plays a dominant role in shaping teenagers' body image perceptions, particularly through its For You Page (FYP), where diet culture, fitness trends, and aesthetic body norms are frequently promoted~\cite{wang2022tiktok}.}


 Nevertheless, the majority of teens' discussions we observed revealed broader and more diverse body image struggles that also took place offline. Therefore, it is important to exercise caution in attributing all body image issues to social media platforms like Instagram as research also attributed these issues to a complex interplay of factors such as pre-existing vulnerabilities, cultural norms, usage patterns, and individual differences \cite{cayla2023impact, gorska2023influence, sagrera2022social}.
 At the same time, we should acknowledge that social media also provides a positive space for adolescents to openly express their personal struggles with negative body image and receive support.
 Prior research suggested that private messages on social media generally lead to a higher likelihood of receiving help, in terms of emotional and information support~\cite{huh2023help}, and higher quality of supportive messages compared to public posts~\cite{liu2018modeling}. In this context, teens in our study sought support in a more empathetic avenue, which helped them share their body image struggles. However, in some cases, the type of help they received helped continue the cycle of self-judgment, as we explain in the next section \ref{5.2}. 
 
Prior research \cite{vandenbosch2022social} has pointed to the importance of leveraging social media as a channel for supporting young people with negative body image. Several social media campaigns such as the body positivity movement, promoting acceptance of all body types and emphasizing health over appearance, have been a key effort in supporting teens dealing with body image concerns \cite{Lazuka2020Are}. However, these efforts have focused primarily on public-facing content, often overlooking the importance of extending support to the more intimate, private conversations where youth are most vulnerable and engage in deeper self-evaluations and appearance-related insecurities \cite{rubio2022youths}. Our research highlights the need for resources and interventions that address these personal interactions. By promoting positive dialogue and peer support \cite{akter2025examining, akter_evaluating_2023} within private messaging and closed groups, social media platforms can empower teens to navigate their body image concerns more effectively. This approach complements public campaigns by ensuring that support extends into all facets of teens' social media experiences, both public and private.

While private exchanges on social media offer essential support for teens, they also act as a double-edged sword by sometimes exposing them to content that perpetuates negative self-perceptions and unhealthy behaviors. Our findings revealed that teens frequently engaged in body shaming discussions or used private messages to share the severe consequences of negative body image, often discussing unhealthy and risky behaviors such as eating disorders.
Prior research has consistently suggested that such disclosures are very common among teens on different social media platforms~\cite{nova2022cultivating, corradini2023dark,pater2019exploring}. 
However, most of these studies did not examine how these issues are discussed privately where the influence of adolescent peers is proven to have an effect~\cite{chung2021adolescent}. Therefore, this underscores the urgent need for social media platforms 
to take a more active role in educating teens about the negative consequences of private interactions and content and their potential negative impacts. Meta's recent initiative to obscure content related to eating disorders for teenage users is a step in the right direction~\cite{Metatohi35:online}. However, we argue against taking similar measurements in censoring private conversations about body image. Instead, social media platforms should consider implementing gentle interventions, like real-time nudges \cite{agha2023strike}, to encourage healthier online interactions among teens, especially in group settings. These nudges are particularly effective in providing proactive guidance that helps prevent risks teens may struggle to address, especially when these risks come from friends rather than strangers. 



\subsection{Encouraging Effective Peer Responses to Body Image Disclosures (RQ2)}
\label{5.2}
We uncovered that peers' responses to body image disclosures varied considerably based on the type of disclosure. One of the most common responses we observed was that peers mostly validated negative body image disclosures, especially when teens judged themselves or sought reassurance. Although this type of response appeared to be helpful at first glance, prior research \cite{mills2017fat, ambwani2017challenging} suggested that it could also play a role in sustaining negative body image discussions. Validating these concerns (e.g., saying "You're not fat, you look great") may not be the most effective response, as it can perpetuate the cycle of negative body image talk without addressing the underlying issues that contribute to these feelings~\cite{mills2019impact}. While we saw teens desired to be reassured about their appearance, this type of validation did not necessarily improve body satisfaction in many occurrences as they often did not believe such reassurances or viewed them as insincere. Instead, challenging these types of talks and emphasizing that appearance is not the most important aspect of a person often leads to less engagement in these talks in the future and leads to better emotional outcomes \cite{ambwani2017challenging}. 

Furthermore, we found that peers often changed the subject to avoid engaging with disclosures about negative body image struggles and self-judgment.
While these responses appeared to be unhelpful or dismissive, they may represent a subtle form of support. By gently steering the conversation away from negative self-focus, peers may help prevent the reinforcement of harmful self-perceptions and discourage teens from engaging in negative body discussions that can reinforce body dissatisfaction and unhealthy body image norms \cite{lawler2011body}. This does not imply that peers should ignore such discussions; rather, acknowledging the individual's feelings and then redirecting the conversation to positive, non-appearance-related qualities can be more effective \cite{mills2019impact}. For example, emphasizing personal strengths, achievements, or values unrelated to physical appearance can promote a more holistic sense of self-worth and reduce the focus on body image \cite{alleva2015expand}. Moreover, modeling and encouraging positive self-talk and body acceptance, as shown in our results, can foster a supportive environment that mitigates the impact of societal pressures on body image. While we observed that youth generally showed positive feedback and supportive responses, some negative reactions did arise, highlighting the need for intentional education and awareness programs to help address these issues effectively.
Prior work suggested that appearance concerns from high social media use can be mitigated by media literacy and appreciation of differences \cite{burnette2017don}. 
It is essential for teens to learn to support self-acceptance in others, rather than offering validation that might unintentionally reinforce negative self-perceptions. 

Additionally, we found in conversations in which teens body-shamed others, peers also mostly criticized others rather than standing up for the targeted individuals. This behavior aligns with social contagion theory \cite{christakis2013social}, which suggests that attitudes and behaviors spread through groups as individuals mimic those around them. In our study, we observed how quickly body-shaming became contagious, as the actions of one or two individuals influenced the entire group’s response. \edit{When negative comments began to circulate, peer influence and social conformity could have reinforced those negative comments as teens perhaps feared exclusion and aligned themselves with group norms rather than challenging them \cite{kamelabad2024conformity}.
At the same time, the bystander effect could have contributed to hesitation, with individuals assuming someone else should intervene, leading to collective passivity~\cite{orlando2020peer, you2019bystander}. Rather than disrupting harmful behaviors, group members often stayed silent or joined in, perhaps further normalizing the criticism. Additionally, ingroup vs. outgroup bias could have made it even less likely for someone to defend the target, as teens likely prioritized maintaining group cohesion over standing up for an outsider~\cite{hogg2016social, kwak2015exploring}. As a result, even those uncomfortable with body-shaming may have hesitated to intervene, fearing social isolation or becoming the next target themselves.
Given how quickly these behaviors can spread, it is crucial to teach adolescents how to navigate these situations as bystanders \cite{difranzo2018upstanding}. Providing them with strategies for positive and assertive intervention can help disrupt the cycle of negativity and foster a safer, more supportive social environment.}



\subsection{Dyadic vs. Group Dynamics in Adolescent Private Body Image Disclosures (RQ3)}
\label{5.3}
We revealed important distinctions in how teens disclose and respond to body image disclosures in one-on-one (dyadic) versus group settings. Consistent with prior research, individuals tend to share less intimate information in larger groups than in dyadic conversations, especially when dealing with sensitive topics like body image \cite{kruger2019individual}. This aligns with the idea that highly personal content is more emotionally charged and involves a greater risk of judgment or misunderstanding, making individuals less likely to disclose such information in group settings \cite{rubio2022youths}. In our study, we observed that dyadic conversations served as safer spaces where teens felt more comfortable sharing personal struggles with negative body image and were more likely to receive supportive feedback. This suggests that promoting private, meaningful relationships with a smaller circle of close friends may be more protective for teens than fostering larger networks \cite{chacon2023student}. Social media platforms could leverage affordances that encourage intimate one-to-one interactions, as they may provide teens with focused, personalized support in lower-risk environments.

In contrast, group settings were associated with higher rates of body-shaming and critical exchanges, likely influenced by social pressures and the diminished accountability that can arise in larger discussions. Prior studies emphasize that peer pressure in group settings can strongly influence adolescent behaviors, pushing individuals to conform either to avoid becoming targets themselves or to fit in \cite{yang2022longitudinal}. Building on this, our findings show that group affordances can inadvertently enable negative dynamics by amplifying criticisms like body-shaming. To mitigate this, platforms could create environments that promote positive group norms and nudge teens toward prosocial behaviors~\cite{agha2023strike}. \edit{Real-time algorithmic nudges could intervene at key moments by detecting harmful language and prompting users to rephrase messages before posting~\cite{alemany2019enhancing, masaki2020exploring}. Additionally, personalized sensitivity filters could allow users to blur or hide negative content, while automated educational pop-ups could provide immediate feedback on group norms and the impact of harmful discourse~\cite{badillo2019stranger}. 
In addition, prior research on adolescent online safety highlights 
the role of resilience in helping teens navigate online risks~\cite{wisniewski2016dear, park2024resilience, d2013cope, badillo2020towards}. 
While platform-based interventions such as sensitivity filters and real-time nudges can help reduce exposure to harmful interactions, they may not completely eliminate negative experiences. In such cases, fostering resilience becomes crucial in equipping teens with the skills to manage these challenges effectively. For those targeted by harmful online interactions, such as body shaming, resilience can promote coping strategies like seeking help, responding proactively, and maintaining digital well-being. At the same time, cultivating resilience among users who might engage in harmful behaviors can encourage self-regulation, awareness of consequences, and responsible online conduct. 
 By integrating resilience-building strategies with real-time intervention mechanisms, such as risk alerts, preventive nudges, and structured guidelines for group engagement, social media platforms can more effectively mitigate peer pressure, prevent harmful interactions, and reinforce positive group norms~\cite{ashktorab2016designing, akter2024towards, garaigordobil2015effects, mcnally2018co}, ultimately fostering a healthier and more supportive online environment.}

\subsection{Implications for Design}
Our study provides insights into how teens engage in discussions about body image on private social media channels with peers, highlighting specific mechanisms that could enhance adolescents' well-being in these digital spaces.

\paragraph{Design Empathy-Based Recommendations for Peer Support}
We found that teens frequently engaged in self-judgment or sought reassurance from their peers during body image discussions, revealing a strong need for support within these conversations. However, peers were not always equipped to provide the appropriate response or lacked the confidence to offer effective support, particularly in group settings where reactions could sometimes turn critical. Research suggests that, without targeted guidance, peers may struggle to develop empathic response skills on their own \cite{sharma2020computational}. To encourage more empathetic and constructive interactions, and in light of the suggestions made earlier in Section \ref{5.2}, social media platforms could introduce context-sensitive support recommendations. These recommendations would guide users in offering thoughtful feedback tailored to the nature of the conversation. These recommendations might include prompts or suggested responses that help teens acknowledge each other’s feelings, offer positive encouragement, or gently redirect negative self-talk. By providing these tools, platforms could empower teens to respond more effectively and create a supportive digital environment that encourages healthy body image discussions.

\paragraph{Encourage Safe Self-Disclosure in One-on-One Settings}
Teens in our study were more likely to share personal and vulnerable feelings about their body image in one-on-one (dyadic) conversations compared to group settings, where they might feel more exposed or judged. To support this behavior, as briefly suggested in section \ref{5.3}, social media platforms could implement private journaling features, enabling users to freely express their thoughts and feelings about body image, much like a personal diary. In addition, the platforms could include prompts or guided questions to encourage deeper self-expression, helping teens articulate their experiences and struggles more clearly.
This feature would provide a dedicated, safe space for self-reflection, reducing concerns about judgment or exposure. Additionally, users could selectively share specific entries with a trusted friend, fostering supportive, private connections that echo the intimacy of dyadic conversations. Research has shown that sharing in these smaller settings can enhance support and encourage openness~\cite{ghosh2020circle}. By mirroring the comfort and security found in one-on-one interactions, this journaling feature could create a valuable tool for teens to address and process body image concerns within a framework of trust and support.

\paragraph{Implement Moderation Tools for Healthy Group Communication}
Group conversations as we saw in our study often become spaces for body-shaming, sometimes masked as humor, where teens may feel uncomfortable or even targeted, making it challenging for them to express themselves openly. To address this, as briefly explained in section \ref{5.1}, platforms could incorporate real-time moderation tools that help guide conversations toward more positive interactions, a feature that teens themselves have expressed a preference for \cite{agha2023strike}. These tools could detect when conversations begin to change direction into harmful territory, such as body-shaming, and issue gentle notifications to prompt a tone shift or remind users of community guidelines. Additionally, social media platforms could also implement a body-shaming detection tool that could monitor for specific phrases, suggesting healthier responses or providing teens the option to exit the conversation or mute specific users. This combination of features would empower teens to distance themselves from damaging discussions, supporting their mental well-being and encouraging a more supportive online environment.

\section{Limitations and Future Work}
There are several limitations of our work that inform future research directions. First, we had demographic details only from participants who donated their data, but not from those they conversed with. Although we could sometimes guess the nature of the relationships, we lacked specific information, such as the age and gender of the conversation partners. While the participants predominantly appeared to be teens, it is possible that some of these conversations involved older adults. 
\edit{Another limitation of this work that could be explored in future research is that we did not comprehensively analyze the temporal variation in these discussions. While we qualitatively examined how the conversations shifted, we did not quantify the extent to which time influenced these changes.}
Furthermore, the dataset we relied on in this study ~\cite{razi2022instagram} was collected solely from Instagram. Therefore, to assess the generalizability of our findings, we recommend that future research explore various social media platforms to identify the unique challenges or protective factors associated with each platform among youth.
In addition, our sample was skewed toward older adolescents. Hence, our findings may not be fully generalizable to teens of different ages. Future research should aim to validate our results with a broader and more diverse youth demographic to ensure their applicability. It would also be beneficial to examine how age influences both body image conversations and the nature of responses within these discussions.

\section{Conclusion}
We investigated teens' body image disclosures in private Instagram conversations and their peers' responses. We found that they frequently received support when disclosing their struggles with negative body image, participated in criticism when engaging in body shaming, and were met with appreciation when promoting positive body image.
The main takeaway of this study is that private social media settings can significantly influence how teens discuss and respond to body image, with private, one-on-one conversations often fostering supportive exchanges, while group settings tend to amplify body-shaming. This difference underscores the potential for social media platforms to enhance positive interactions by promoting private one-on-one discussions for sensitive topics and introducing prosocial features in group settings. Such changes could help create a healthier and more supportive online environment for teens dealing with body image concerns.

\begin{acks}
This research is supported in part by the U.S. National Science Foundation under grants \#IIP-2329976, \#IIS-2333207 and by the William T. Grant Foundation grant \#187941. Any opinions, findings, and conclusions or recommendations expressed in this material are those of the authors and do not necessarily reflect the views of the research sponsors. We would also like to thank Nafees-ul Haque and Sarvech Qadir for their support with relevancy coding and data retrieval. We also extend our gratitude to all the participants who donated their data and contributed towards our research.
\end{acks}

\bibliographystyle{ACM-Reference-Format}
\bibliography{07-References}


\begin{thebibliography}{109}


\ifx \showCODEN    \undefined \def \showCODEN     #1{\unskip}     \fi
\ifx \showDOI      \undefined \def \showDOI       #1{#1}\fi
\ifx \showISBNx    \undefined \def \showISBNx     #1{\unskip}     \fi
\ifx \showISBNxiii \undefined \def \showISBNxiii  #1{\unskip}     \fi
\ifx \showISSN     \undefined \def \showISSN      #1{\unskip}     \fi
\ifx \showLCCN     \undefined \def \showLCCN      #1{\unskip}     \fi
\ifx \shownote     \undefined \def \shownote      #1{#1}          \fi
\ifx \showarticletitle \undefined \def \showarticletitle #1{#1}   \fi
\ifx \showURL      \undefined \def \showURL       {\relax}        \fi
\providecommand\bibfield[2]{#2}
\providecommand\bibinfo[2]{#2}
\providecommand\natexlab[1]{#1}
\providecommand\showeprint[2][]{arXiv:#2}

\bibitem[Agha et~al\mbox{.}(2023)]%
        {agha2023strike}
\bibfield{author}{\bibinfo{person}{Zainab Agha}, \bibinfo{person}{Karla Badillo-Urquiola}, {and} \bibinfo{person}{Pamela~J Wisniewski}.} \bibinfo{year}{2023}\natexlab{}.
\newblock \showarticletitle{" Strike at the Root": Co-designing Real-Time Social Media Interventions for Adolescent Online Risk Prevention}.
\newblock \bibinfo{journal}{\emph{Proceedings of the ACM on Human-Computer Interaction}} \bibinfo{volume}{7}, \bibinfo{number}{CSCW1} (\bibinfo{year}{2023}), \bibinfo{pages}{1--32}.
\newblock


\bibitem[Ajmal et~al\mbox{.}(2019)]%
        {ajmal2019impact}
\bibfield{author}{\bibinfo{person}{Amna Ajmal} {et~al\mbox{.}}} \bibinfo{year}{2019}\natexlab{}.
\newblock \showarticletitle{The impact of body image on self-esteem in adolescents}.
\newblock \bibinfo{journal}{\emph{Clinical and Counselling Psychology Review}} \bibinfo{volume}{1}, \bibinfo{number}{1} (\bibinfo{year}{2019}), \bibinfo{pages}{44--54}.
\newblock


\bibitem[Akter et~al\mbox{.}(2024)]%
        {akter2024towards}
\bibfield{author}{\bibinfo{person}{Mamtaj Akter}, \bibinfo{person}{Zainab Agha}, \bibinfo{person}{Ashwaq Alsoubai}, \bibinfo{person}{Naima Ali}, {and} \bibinfo{person}{Pamela Wisniewski}.} \bibinfo{year}{2024}\natexlab{}.
\newblock \showarticletitle{Towards Collaborative Family-Centered Design for Online Safety, Privacy and Security}. In \bibinfo{booktitle}{\emph{Akter, M., Agha, Z., Alsoubai, A., Ali, N. S., Wisniewski, P., (2024)“Towards Collaborative Family-Centered Design for Online Safety, Privacy and Security” Extended Abstract presented at the ACM Conference on Human Factors in Computing Systems Workshop on Family-Centered Design, (CHI 2024)}}.
\newblock


\bibitem[Akter et~al\mbox{.}(2025)]%
        {akter2025examining}
\bibfield{author}{\bibinfo{person}{Mamtaj Akter}, \bibinfo{person}{Jess Kropczynski}, \bibinfo{person}{Heather Lipford}, {and} \bibinfo{person}{Pamela Wisniewski}.} \bibinfo{year}{2025}\natexlab{}.
\newblock \showarticletitle{Examining Caregiving Roles to Differentiate the Effects of Using a Mobile App for Community Oversight for Privacy and Security}.
\newblock \bibinfo{journal}{\emph{Proceedings on Privacy Enhancing Technologies}} (\bibinfo{year}{2025}).
\newblock


\bibitem[Akter et~al\mbox{.}(2023)]%
        {akter_evaluating_2023}
\bibfield{author}{\bibinfo{person}{Mamtaj Akter}, \bibinfo{person}{Madiha Tabassum}, \bibinfo{person}{Nazmus~Sakib Miazi}, \bibinfo{person}{Leena Alghamdi}, \bibinfo{person}{Jess Kropczynski}, \bibinfo{person}{Pamela~J. Wisniewski}, {and} \bibinfo{person}{Heather Lipford}.} \bibinfo{year}{2023}\natexlab{}.
\newblock \showarticletitle{Evaluating the Impact of Community Oversight for Managing Mobile Privacy and Security}. In \bibinfo{booktitle}{\emph{Nineteenth Symposium on Usable Privacy and Security (SOUPS 2023)}}. \bibinfo{publisher}{USENIX Association}, \bibinfo{address}{Anaheim, CA}, \bibinfo{pages}{437--456}.
\newblock
\showISBNx{978-1-939133-36-6}
\urldef\tempurl%
\url{https://www.usenix.org/conference/soups2023/presentation/akter}
\showURL{%
\tempurl}


\bibitem[Alemany et~al\mbox{.}(2019)]%
        {alemany2019enhancing}
\bibfield{author}{\bibinfo{person}{Jos{\'e} Alemany}, \bibinfo{person}{Elena Del~Val}, \bibinfo{person}{Juan Alberola}, {and} \bibinfo{person}{Ana Garc{\'\i}a-Fornes}.} \bibinfo{year}{2019}\natexlab{}.
\newblock \showarticletitle{Enhancing the privacy risk awareness of teenagers in online social networks through soft-paternalism mechanisms}.
\newblock \bibinfo{journal}{\emph{International Journal of Human-Computer Studies}}  \bibinfo{volume}{129} (\bibinfo{year}{2019}), \bibinfo{pages}{27--40}.
\newblock


\bibitem[Alleva et~al\mbox{.}(2015)]%
        {alleva2015expand}
\bibfield{author}{\bibinfo{person}{Jessica~M Alleva}, \bibinfo{person}{Carolien Martijn}, \bibinfo{person}{Gerard~JP Van~Breukelen}, \bibinfo{person}{Anita Jansen}, {and} \bibinfo{person}{Kai Karos}.} \bibinfo{year}{2015}\natexlab{}.
\newblock \showarticletitle{Expand Your Horizon: A programme that improves body image and reduces self-objectification by training women to focus on body functionality}.
\newblock \bibinfo{journal}{\emph{Body image}}  \bibinfo{volume}{15} (\bibinfo{year}{2015}), \bibinfo{pages}{81--89}.
\newblock


\bibitem[Alluhidan et~al\mbox{.}(2024)]%
        {alluhidan2024teen}
\bibfield{author}{\bibinfo{person}{Abdulmalik Alluhidan}, \bibinfo{person}{Mamtaj Akter}, \bibinfo{person}{Ashwaq Alsoubai}, \bibinfo{person}{Jinkyung~Katie Park}, {and} \bibinfo{person}{Pamela Wisniewski}.} \bibinfo{year}{2024}\natexlab{}.
\newblock \showarticletitle{Teen Talk: The Good, the Bad, and the Neutral of Adolescent Social Media Use}.
\newblock \bibinfo{journal}{\emph{Proceedings of the ACM on Human-Computer Interaction}} \bibinfo{volume}{8}, \bibinfo{number}{CSCW2} (\bibinfo{year}{2024}), \bibinfo{pages}{1--35}.
\newblock


\bibitem[Ambwani et~al\mbox{.}(2017)]%
        {ambwani2017challenging}
\bibfield{author}{\bibinfo{person}{Suman Ambwani}, \bibinfo{person}{Megan Baumgardner}, \bibinfo{person}{Cai Guo}, \bibinfo{person}{Lea Simms}, {and} \bibinfo{person}{Emily Abromowitz}.} \bibinfo{year}{2017}\natexlab{}.
\newblock \showarticletitle{Challenging fat talk: An experimental investigation of reactions to body disparaging conversations}.
\newblock \bibinfo{journal}{\emph{Body Image}}  \bibinfo{volume}{23} (\bibinfo{year}{2017}), \bibinfo{pages}{85--92}.
\newblock


\bibitem[Anixiadis et~al\mbox{.}(2019)]%
        {anixiadis2019effects}
\bibfield{author}{\bibinfo{person}{Fay Anixiadis}, \bibinfo{person}{Eleanor~H Wertheim}, \bibinfo{person}{Rachel Rodgers}, {and} \bibinfo{person}{Brigitte Caruana}.} \bibinfo{year}{2019}\natexlab{}.
\newblock \showarticletitle{Effects of thin-ideal instagram images: The roles of appearance comparisons, internalization of the thin ideal and critical media processing}.
\newblock \bibinfo{journal}{\emph{Body image}}  \bibinfo{volume}{31} (\bibinfo{year}{2019}), \bibinfo{pages}{181--190}.
\newblock


\bibitem[Ashktorab and Vitak(2016)]%
        {ashktorab2016designing}
\bibfield{author}{\bibinfo{person}{Zahra Ashktorab} {and} \bibinfo{person}{Jessica Vitak}.} \bibinfo{year}{2016}\natexlab{}.
\newblock \showarticletitle{Designing cyberbullying mitigation and prevention solutions through participatory design with teenagers}. In \bibinfo{booktitle}{\emph{Proceedings of the 2016 CHI conference on human factors in computing systems}}. \bibinfo{pages}{3895--3905}.
\newblock


\bibitem[Badillo-Urquiola et~al\mbox{.}(2020)]%
        {badillo2020towards}
\bibfield{author}{\bibinfo{person}{Karla Badillo-Urquiola}, \bibinfo{person}{Zainab Agha}, \bibinfo{person}{Mamtaj Akter}, {and} \bibinfo{person}{Pamela Wisniewski}.} \bibinfo{year}{2020}\natexlab{}.
\newblock \showarticletitle{Towards Assets-based Approaches for Adolescent Online Safety}. In \bibinfo{booktitle}{\emph{Badillo-Urquiola, Agha, Z., Akter, K., Wisniewski, P.,(2020)“Towards Assets-Based Approaches for Adolescent Online Safety” Extended Abstract presented at the ACM Conference on Computer-Supported Cooperative Work Workshop on Operationalizing an Assets-Based Design of Technology,(CSCW 2020)}}.
\newblock


\bibitem[Badillo-Urquiola et~al\mbox{.}(2019)]%
        {badillo2019stranger}
\bibfield{author}{\bibinfo{person}{Karla Badillo-Urquiola}, \bibinfo{person}{Diva Smriti}, \bibinfo{person}{Brenna McNally}, \bibinfo{person}{Evan Golub}, \bibinfo{person}{Elizabeth Bonsignore}, {and} \bibinfo{person}{Pamela~J Wisniewski}.} \bibinfo{year}{2019}\natexlab{}.
\newblock \showarticletitle{Stranger danger! social media app features co-designed with children to keep them safe online}. In \bibinfo{booktitle}{\emph{Proceedings of the 18th ACM international conference on interaction design and children}}. \bibinfo{pages}{394--406}.
\newblock


\bibitem[Bonner et~al\mbox{.}(2023)]%
        {bonner2023filters}
\bibfield{author}{\bibinfo{person}{Jolie Bonner}, \bibinfo{person}{Florian Mathis}, \bibinfo{person}{Joseph O'Hagan}, {and} \bibinfo{person}{Mark Mcgill}.} \bibinfo{year}{2023}\natexlab{}.
\newblock \showarticletitle{When filters escape the smartphone: Exploring acceptance and concerns regarding augmented expression of social identity for everyday AR}. In \bibinfo{booktitle}{\emph{Proceedings of the 29th ACM Symposium on Virtual Reality Software and Technology}}. \bibinfo{pages}{1--14}.
\newblock


\bibitem[Braun and Clarke(2012)]%
        {braun2012thematic}
\bibfield{author}{\bibinfo{person}{Virginia Braun} {and} \bibinfo{person}{Victoria Clarke}.} \bibinfo{year}{2012}\natexlab{}.
\newblock \bibinfo{booktitle}{\emph{Thematic analysis.}}
\newblock \bibinfo{publisher}{American Psychological Association}.
\newblock


\bibitem[Brown and Tiggemann(2020)]%
        {brown2020picture}
\bibfield{author}{\bibinfo{person}{Zoe Brown} {and} \bibinfo{person}{Marika Tiggemann}.} \bibinfo{year}{2020}\natexlab{}.
\newblock \showarticletitle{A picture is worth a thousand words: The effect of viewing celebrity Instagram images with disclaimer and body positive captions on women’s body image}.
\newblock \bibinfo{journal}{\emph{Body image}}  \bibinfo{volume}{33} (\bibinfo{year}{2020}), \bibinfo{pages}{190--198}.
\newblock


\bibitem[Burnette et~al\mbox{.}(2017)]%
        {burnette2017don}
\bibfield{author}{\bibinfo{person}{C~Blair Burnette}, \bibinfo{person}{Melissa~A Kwitowski}, {and} \bibinfo{person}{Suzanne~E Mazzeo}.} \bibinfo{year}{2017}\natexlab{}.
\newblock \showarticletitle{“I don’t need people to tell me I’m pretty on social media:” A qualitative study of social media and body image in early adolescent girls}.
\newblock \bibinfo{journal}{\emph{Body Image}}  \bibinfo{volume}{23} (\bibinfo{year}{2017}), \bibinfo{pages}{114--125}.
\newblock


\bibitem[Cayla et~al\mbox{.}(2023)]%
        {cayla2023impact}
\bibfield{author}{\bibinfo{person}{Kirrily Cayla}, \bibinfo{person}{Maitland Eithan}, {and} \bibinfo{person}{Christy Macie}.} \bibinfo{year}{2023}\natexlab{}.
\newblock \showarticletitle{The Impact of Social Media on Body Image and Self-Esteem: A Comparative Study of Adolescent Girls in United States and South Korea}.
\newblock \bibinfo{journal}{\emph{Studies in Social Science \& Humanities}} \bibinfo{volume}{2}, \bibinfo{number}{11} (\bibinfo{year}{2023}), \bibinfo{pages}{30--37}.
\newblock


\bibitem[Chacon et~al\mbox{.}(2023)]%
        {chacon2023student}
\bibfield{author}{\bibinfo{person}{Marco Chacon}, \bibinfo{person}{Rebecca~S Levine}, {and} \bibinfo{person}{Amy Bintliff}.} \bibinfo{year}{2023}\natexlab{}.
\newblock \showarticletitle{Student perceptions: How virtual student-led talking circles promote engagement, social connectedness, and academic benefit}.
\newblock \bibinfo{journal}{\emph{Active Learning in Higher Education}} (\bibinfo{year}{2023}), \bibinfo{pages}{14697874231179238}.
\newblock


\bibitem[Chen et~al\mbox{.}(2021)]%
        {chen2021scaffolding}
\bibfield{author}{\bibinfo{person}{Tianying Chen}, \bibinfo{person}{Kristy Zhang}, \bibinfo{person}{Robert~E Kraut}, {and} \bibinfo{person}{Laura Dabbish}.} \bibinfo{year}{2021}\natexlab{}.
\newblock \showarticletitle{Scaffolding the online peer-support experience: novice supporters' strategies and challenges}.
\newblock \bibinfo{journal}{\emph{Proceedings of the ACM on Human-Computer Interaction}} \bibinfo{volume}{5}, \bibinfo{number}{CSCW2} (\bibinfo{year}{2021}), \bibinfo{pages}{1--30}.
\newblock


\bibitem[Chiu and Yuan(2021)]%
        {chiu2021last}
\bibfield{author}{\bibinfo{person}{Hsuen~Chi Chiu} {and} \bibinfo{person}{Chien~Wen Yuan}.} \bibinfo{year}{2021}\natexlab{}.
\newblock \showarticletitle{To Last Long or to Fade Away: Investigating Users' Instagram Post and Story Practices}. In \bibinfo{booktitle}{\emph{Companion Publication of the 2021 Conference on Computer Supported Cooperative Work and Social Computing}}. \bibinfo{pages}{32--35}.
\newblock


\bibitem[Choukas-Bradley et~al\mbox{.}(2022)]%
        {choukas2022perfect}
\bibfield{author}{\bibinfo{person}{Sophia Choukas-Bradley}, \bibinfo{person}{Savannah~R Roberts}, \bibinfo{person}{Anne~J Maheux}, {and} \bibinfo{person}{Jacqueline Nesi}.} \bibinfo{year}{2022}\natexlab{}.
\newblock \showarticletitle{The perfect storm: A developmental--sociocultural framework for the role of social media in adolescent girls’ body image concerns and mental health}.
\newblock \bibinfo{journal}{\emph{Clinical child and family psychology review}} \bibinfo{volume}{25}, \bibinfo{number}{4} (\bibinfo{year}{2022}), \bibinfo{pages}{681--701}.
\newblock


\bibitem[Christakis and Fowler(2013)]%
        {christakis2013social}
\bibfield{author}{\bibinfo{person}{Nicholas~A Christakis} {and} \bibinfo{person}{James~H Fowler}.} \bibinfo{year}{2013}\natexlab{}.
\newblock \showarticletitle{Social contagion theory: examining dynamic social networks and human behavior}.
\newblock \bibinfo{journal}{\emph{Statistics in medicine}} \bibinfo{volume}{32}, \bibinfo{number}{4} (\bibinfo{year}{2013}), \bibinfo{pages}{556--577}.
\newblock


\bibitem[Chung et~al\mbox{.}(2021)]%
        {chung2021adolescent}
\bibfield{author}{\bibinfo{person}{Alicia Chung}, \bibinfo{person}{Dorice Vieira}, \bibinfo{person}{Tiffany Donley}, \bibinfo{person}{Nicholas Tan}, \bibinfo{person}{Girardin Jean-Louis}, \bibinfo{person}{Kathleen~Kiely Gouley}, \bibinfo{person}{Azizi Seixas}, {et~al\mbox{.}}} \bibinfo{year}{2021}\natexlab{}.
\newblock \showarticletitle{Adolescent peer influence on eating behaviors via social media: scoping review}.
\newblock \bibinfo{journal}{\emph{Journal of medical Internet research}} \bibinfo{volume}{23}, \bibinfo{number}{6} (\bibinfo{year}{2021}), \bibinfo{pages}{e19697}.
\newblock


\bibitem[Cohen et~al\mbox{.}(2019)]%
        {Cohen2019BoPo}
\bibfield{author}{\bibinfo{person}{Rachel Cohen}, \bibinfo{person}{J. Fardouly}, \bibinfo{person}{T. Newton-John}, {and} \bibinfo{person}{A. Slater}.} \bibinfo{year}{2019}\natexlab{}.
\newblock \showarticletitle{\#BoPo on Instagram: An experimental investigation of the effects of viewing body positive content on young women’s mood and body image}.
\newblock \bibinfo{journal}{\emph{New Media \& Society}}  \bibinfo{volume}{21} (\bibinfo{year}{2019}), \bibinfo{pages}{1546 -- 1564}.
\newblock
\urldef\tempurl%
\url{https://doi.org/10.1177/1461444819826530}
\showDOI{\tempurl}


\bibitem[Commission et~al\mbox{.}(2013)]%
        {federal2013children}
\bibfield{author}{\bibinfo{person}{Federal~Trade Commission} {et~al\mbox{.}}} \bibinfo{year}{2013}\natexlab{}.
\newblock \showarticletitle{Children’s online privacy protection rule}.
\newblock \bibinfo{journal}{\emph{COPPA"),"[Online]. Available: https://www. ftc. gov/enforcement/rules/rulemaking-regulatory-reform-proceedings/childrens-online-privacyprotection-rule.[Accessed 23 May 2017]}} (\bibinfo{year}{2013}).
\newblock


\bibitem[Cooney et~al\mbox{.}(2020)]%
        {cooney2020many}
\bibfield{author}{\bibinfo{person}{Gus Cooney}, \bibinfo{person}{Adam~M Mastroianni}, \bibinfo{person}{Nicole Abi-Esber}, {and} \bibinfo{person}{Alison~Wood Brooks}.} \bibinfo{year}{2020}\natexlab{}.
\newblock \showarticletitle{The many minds problem: disclosure in dyadic versus group conversation}.
\newblock \bibinfo{journal}{\emph{Current Opinion in Psychology}}  \bibinfo{volume}{31} (\bibinfo{year}{2020}), \bibinfo{pages}{22--27}.
\newblock


\bibitem[Corradini(2023)]%
        {corradini2023dark}
\bibfield{author}{\bibinfo{person}{Enrico Corradini}.} \bibinfo{year}{2023}\natexlab{}.
\newblock \showarticletitle{The dark threads that weave the web of shame: A network science-inspired analysis of body shaming on Reddit}.
\newblock \bibinfo{journal}{\emph{Information}} \bibinfo{volume}{14}, \bibinfo{number}{8} (\bibinfo{year}{2023}), \bibinfo{pages}{436}.
\newblock


\bibitem[De~Choudhury and De(2014)]%
        {de2014mental}
\bibfield{author}{\bibinfo{person}{Munmun De~Choudhury} {and} \bibinfo{person}{Sushovan De}.} \bibinfo{year}{2014}\natexlab{}.
\newblock \showarticletitle{Mental health discourse on reddit: Self-disclosure, social support, and anonymity}. In \bibinfo{booktitle}{\emph{Proceedings of the international AAAI conference on web and social media}}, Vol.~\bibinfo{volume}{8}. \bibinfo{pages}{71--80}.
\newblock


\bibitem[Deogracias(2015)]%
        {deogracias2015danah}
\bibfield{author}{\bibinfo{person}{Alicia Deogracias}.} \bibinfo{year}{2015}\natexlab{}.
\newblock \bibinfo{title}{Danah Boyd: It’s Complicated: The Social Lives of Networked Teens: Yale University Press, New Haven, Connecticut, 2014, pp. 296, ISBN 973-0-300-16631-6}.
\newblock
\newblock


\bibitem[Devakumar et~al\mbox{.}(2021)]%
        {devakumar2021review}
\bibfield{author}{\bibinfo{person}{Anjali Devakumar}, \bibinfo{person}{Jay Modh}, \bibinfo{person}{Bahador Saket}, \bibinfo{person}{Eric~PS Baumer}, {and} \bibinfo{person}{Munmun De~Choudhury}.} \bibinfo{year}{2021}\natexlab{}.
\newblock \showarticletitle{A review on strategies for data collection, reflection, and communication in eating disorder apps}. In \bibinfo{booktitle}{\emph{Proceedings of the 2021 CHI conference on human factors in computing systems}}. \bibinfo{pages}{1--19}.
\newblock


\bibitem[DeVito et~al\mbox{.}(2017)]%
        {devito2017platforms}
\bibfield{author}{\bibinfo{person}{Michael~A DeVito}, \bibinfo{person}{Jeremy Birnholtz}, {and} \bibinfo{person}{Jeffery~T Hancock}.} \bibinfo{year}{2017}\natexlab{}.
\newblock \showarticletitle{Platforms, people, and perception: Using affordances to understand self-presentation on social media}. In \bibinfo{booktitle}{\emph{Proceedings of the 2017 ACM conference on computer supported cooperative work and social computing}}. \bibinfo{pages}{740--754}.
\newblock


\bibitem[d'Haenens et~al\mbox{.}(2013)]%
        {d2013cope}
\bibfield{author}{\bibinfo{person}{Leen d'Haenens}, \bibinfo{person}{Sofie Vandoninck}, {and} \bibinfo{person}{Ver{\'o}nica Donoso}.} \bibinfo{year}{2013}\natexlab{}.
\newblock \showarticletitle{How to cope and build online resilience?}
\newblock  (\bibinfo{year}{2013}).
\newblock


\bibitem[DiFranzo et~al\mbox{.}(2018)]%
        {difranzo2018upstanding}
\bibfield{author}{\bibinfo{person}{Dominic DiFranzo}, \bibinfo{person}{Samuel~Hardman Taylor}, \bibinfo{person}{Franccesca Kazerooni}, \bibinfo{person}{Olivia~D Wherry}, {and} \bibinfo{person}{Natalya~N Bazarova}.} \bibinfo{year}{2018}\natexlab{}.
\newblock \showarticletitle{Upstanding by design: Bystander intervention in cyberbullying}. In \bibinfo{booktitle}{\emph{Proceedings of the 2018 CHI conference on human factors in computing systems}}. \bibinfo{pages}{1--12}.
\newblock


\bibitem[Ding et~al\mbox{.}(2023)]%
        {ding2023infrastructural}
\bibfield{author}{\bibinfo{person}{Xianghua Ding}, \bibinfo{person}{Linda Tran}, \bibinfo{person}{Yanling Liu}, \bibinfo{person}{Conor O'Neill}, {and} \bibinfo{person}{Stephen Lindsay}.} \bibinfo{year}{2023}\natexlab{}.
\newblock \showarticletitle{Infrastructural Work Behind The Scene: A Study of Formalized Peer-support Practices for Mental Health}. In \bibinfo{booktitle}{\emph{Proceedings of the 2023 CHI Conference on Human Factors in Computing Systems}}. \bibinfo{pages}{1--14}.
\newblock


\bibitem[Dion et~al\mbox{.}(2015)]%
        {dion2015development}
\bibfield{author}{\bibinfo{person}{Jacinthe Dion}, \bibinfo{person}{Marie-Eve Blackburn}, \bibinfo{person}{Julie Auclair}, \bibinfo{person}{Luc Laberge}, \bibinfo{person}{Suzanne Veillette}, \bibinfo{person}{Marco Gaudreault}, \bibinfo{person}{Patrick Vachon}, \bibinfo{person}{Michel Perron}, {and} \bibinfo{person}{Evelyne Touchette}.} \bibinfo{year}{2015}\natexlab{}.
\newblock \showarticletitle{Development and aetiology of body dissatisfaction in adolescent boys and girls}.
\newblock \bibinfo{journal}{\emph{International journal of adolescence and youth}} \bibinfo{volume}{20}, \bibinfo{number}{2} (\bibinfo{year}{2015}), \bibinfo{pages}{151--166}.
\newblock


\bibitem[Drag(1970)]%
        {drag1970experimenter}
\bibfield{author}{\bibinfo{person}{Richard~Michael Drag}.} \bibinfo{year}{1970}\natexlab{}.
\newblock \showarticletitle{Experimenter behavior and group size as variables influencing self-disclosure.}
\newblock  (\bibinfo{year}{1970}).
\newblock


\bibitem[Eagle and Ringland(2023)]%
        {eagle2023you}
\bibfield{author}{\bibinfo{person}{Tessa Eagle} {and} \bibinfo{person}{Kathryn~E Ringland}.} \bibinfo{year}{2023}\natexlab{}.
\newblock \showarticletitle{“You Can't Possibly Have ADHD”: Exploring Validation and Tensions around Diagnosis within Unbounded ADHD Social Media Communities}. In \bibinfo{booktitle}{\emph{Proceedings of the 25th International ACM SIGACCESS Conference on Computers and Accessibility}}. \bibinfo{pages}{1--17}.
\newblock


\bibitem[ebecca Klar(2024)]%
        {Metatohi35:online}
\bibfield{author}{\bibinfo{person}{ebecca Klar}.} \bibinfo{year}{2024}\natexlab{}.
\newblock \bibinfo{title}{Meta to hide self-harm, eating disorder content from teen users  | CBS 42}.
\newblock \bibinfo{howpublished}{\url{https://www.cbs42.com/news/national/meta-to-hide-self-harm-eating-disorder-content-from-teen-users/}}.
\newblock
\newblock
\shownote{(Accessed on 01/25/2024)}.


\bibitem[Faelens et~al\mbox{.}(2021)]%
        {faelens2021relationship}
\bibfield{author}{\bibinfo{person}{Lien Faelens}, \bibinfo{person}{Kristof Hoorelbeke}, \bibinfo{person}{Ruben Cambier}, \bibinfo{person}{Jill Van~Put}, \bibinfo{person}{Eowyn Van~de Putte}, \bibinfo{person}{Rudi De~Raedt}, {and} \bibinfo{person}{Ernst~HW Koster}.} \bibinfo{year}{2021}\natexlab{}.
\newblock \showarticletitle{The relationship between Instagram use and indicators of mental health: A systematic review}.
\newblock \bibinfo{journal}{\emph{Computers in Human Behavior Reports}}  \bibinfo{volume}{4} (\bibinfo{year}{2021}), \bibinfo{pages}{100121}.
\newblock


\bibitem[Faraj and Azad(2012)]%
        {faraj2012materiality}
\bibfield{author}{\bibinfo{person}{Samer Faraj} {and} \bibinfo{person}{Bijan Azad}.} \bibinfo{year}{2012}\natexlab{}.
\newblock \showarticletitle{The materiality of technology: An affordance perspective}.
\newblock \bibinfo{journal}{\emph{Materiality and organizing: Social interaction in a technological world}} \bibinfo{volume}{237}, \bibinfo{number}{1} (\bibinfo{year}{2012}), \bibinfo{pages}{237--258}.
\newblock


\bibitem[Fardouly and Rapee(2019)]%
        {fardouly2019impact}
\bibfield{author}{\bibinfo{person}{Jasmine Fardouly} {and} \bibinfo{person}{Ronald~M Rapee}.} \bibinfo{year}{2019}\natexlab{}.
\newblock \showarticletitle{The impact of no-makeup selfies on young women’s body image}.
\newblock \bibinfo{journal}{\emph{Body image}}  \bibinfo{volume}{28} (\bibinfo{year}{2019}), \bibinfo{pages}{128--134}.
\newblock


\bibitem[Fatt and Fardouly(2023)]%
        {fatt2023digital}
\bibfield{author}{\bibinfo{person}{Scott~J Fatt} {and} \bibinfo{person}{Jasmine Fardouly}.} \bibinfo{year}{2023}\natexlab{}.
\newblock \showarticletitle{Digital social evaluation: Relationships between receiving likes, comments, and follows on social media and adolescents’ body image concerns}.
\newblock \bibinfo{journal}{\emph{Body Image}}  \bibinfo{volume}{47} (\bibinfo{year}{2023}), \bibinfo{pages}{101621}.
\newblock


\bibitem[Feng et~al\mbox{.}(2024)]%
        {feng2024mapping}
\bibfield{author}{\bibinfo{person}{KJ~Kevin Feng}, \bibinfo{person}{Xander Koo}, \bibinfo{person}{Lawrence Tan}, \bibinfo{person}{Amy Bruckman}, \bibinfo{person}{David~W McDonald}, {and} \bibinfo{person}{Amy~X Zhang}.} \bibinfo{year}{2024}\natexlab{}.
\newblock \showarticletitle{Mapping the Design Space of Teachable Social Media Feed Experiences}. In \bibinfo{booktitle}{\emph{Proceedings of the CHI Conference on Human Factors in Computing Systems}}. \bibinfo{pages}{1--20}.
\newblock


\bibitem[Fettach and Benhiba(2019)]%
        {fettach2019pro}
\bibfield{author}{\bibinfo{person}{Yousra Fettach} {and} \bibinfo{person}{Lamia Benhiba}.} \bibinfo{year}{2019}\natexlab{}.
\newblock \showarticletitle{Pro-eating disorders and pro-recovery communities on Reddit: text and network comparative analyses}. In \bibinfo{booktitle}{\emph{Proceedings of the 21st International Conference on Information Integration and Web-Based Applications \& Services}}. \bibinfo{pages}{277--286}.
\newblock


\bibitem[Fleming and Darley(1991)]%
        {fleming1991mixed}
\bibfield{author}{\bibinfo{person}{John~H Fleming} {and} \bibinfo{person}{John~M Darley}.} \bibinfo{year}{1991}\natexlab{}.
\newblock \showarticletitle{Mixed messages: The multiple audience problem and strategic communication}.
\newblock \bibinfo{journal}{\emph{Social cognition}} \bibinfo{volume}{9}, \bibinfo{number}{1} (\bibinfo{year}{1991}), \bibinfo{pages}{25--46}.
\newblock


\bibitem[Garaigordobil and Mart{\'\i}nez-Valderrey(2015)]%
        {garaigordobil2015effects}
\bibfield{author}{\bibinfo{person}{Maite Garaigordobil} {and} \bibinfo{person}{Vanesa Mart{\'\i}nez-Valderrey}.} \bibinfo{year}{2015}\natexlab{}.
\newblock \showarticletitle{Effects of Cyberprogram 2.0 on" face-to-face" bullying, cyberbullying, and empathy}.
\newblock \bibinfo{journal}{\emph{Psicothema}} (\bibinfo{year}{2015}), \bibinfo{pages}{45--51}.
\newblock


\bibitem[Ghosh et~al\mbox{.}(2020)]%
        {ghosh2020circle}
\bibfield{author}{\bibinfo{person}{Arup~Kumar Ghosh}, \bibinfo{person}{Charles~E Hughes}, {and} \bibinfo{person}{Pamela~J Wisniewski}.} \bibinfo{year}{2020}\natexlab{}.
\newblock \showarticletitle{Circle of trust: a new approach to mobile online safety for families}. In \bibinfo{booktitle}{\emph{Proceedings of the 2020 CHI Conference on Human Factors in Computing Systems}}. \bibinfo{pages}{1--14}.
\newblock


\bibitem[Gibson and Trnka(2020)]%
        {gibson2020young}
\bibfield{author}{\bibinfo{person}{Kerry Gibson} {and} \bibinfo{person}{Susanna Trnka}.} \bibinfo{year}{2020}\natexlab{}.
\newblock \showarticletitle{Young people's priorities for support on social media:“It takes trust to talk about these issues”}.
\newblock \bibinfo{journal}{\emph{Computers in human behavior}}  \bibinfo{volume}{102} (\bibinfo{year}{2020}), \bibinfo{pages}{238--247}.
\newblock


\bibitem[Goray and Schoenebeck(2022)]%
        {goray2022youths}
\bibfield{author}{\bibinfo{person}{Cami Goray} {and} \bibinfo{person}{Sarita Schoenebeck}.} \bibinfo{year}{2022}\natexlab{}.
\newblock \showarticletitle{Youths' Perceptions of Data Collection in Online Advertising and Social Media}.
\newblock \bibinfo{journal}{\emph{Proceedings of the ACM on Human-Computer Interaction}} \bibinfo{volume}{6}, \bibinfo{number}{CSCW2} (\bibinfo{year}{2022}), \bibinfo{pages}{1--27}.
\newblock


\bibitem[G{\'o}rska et~al\mbox{.}(2023)]%
        {gorska2023influence}
\bibfield{author}{\bibinfo{person}{Dominika G{\'o}rska}, \bibinfo{person}{Kamila {\'S}wiercz}, \bibinfo{person}{Magdalena Majcher}, \bibinfo{person}{Ma{\l}gorzata Sierpie{\'n}}, \bibinfo{person}{Monika Majcher}, \bibinfo{person}{Agata Pikulicka}, \bibinfo{person}{Aleksandra Karwa{\'n}ska}, \bibinfo{person}{Aleksandra Kulbat}, \bibinfo{person}{Piotr Brzychczy}, {and} \bibinfo{person}{Mateusz Kulbat}.} \bibinfo{year}{2023}\natexlab{}.
\newblock \showarticletitle{The Influence of social media on developing body image dissatisfaction and eating disorders}.
\newblock \bibinfo{journal}{\emph{Journal of Education, Health and Sport}} \bibinfo{volume}{22}, \bibinfo{number}{1} (\bibinfo{year}{2023}), \bibinfo{pages}{56--62}.
\newblock


\bibitem[Grogan(2021)]%
        {grogan2021body}
\bibfield{author}{\bibinfo{person}{Sarah Grogan}.} \bibinfo{year}{2021}\natexlab{}.
\newblock \bibinfo{booktitle}{\emph{Body image: Understanding body dissatisfaction in men, women and children}}.
\newblock \bibinfo{publisher}{Routledge}.
\newblock


\bibitem[Hadden et~al\mbox{.}(2018)]%
        {hadden2018readability}
\bibfield{author}{\bibinfo{person}{Kristie~B Hadden}, \bibinfo{person}{Latrina Prince}, \bibinfo{person}{Laura James}, \bibinfo{person}{Jennifer Holland}, {and} \bibinfo{person}{Christopher~R Trudeau}.} \bibinfo{year}{2018}\natexlab{}.
\newblock \showarticletitle{Readability of human subjects training materials for research}.
\newblock \bibinfo{journal}{\emph{Journal of Empirical Research on Human Research Ethics}} \bibinfo{volume}{13}, \bibinfo{number}{1} (\bibinfo{year}{2018}), \bibinfo{pages}{95--100}.
\newblock


\bibitem[Hogg(2016)]%
        {hogg2016social}
\bibfield{author}{\bibinfo{person}{Michael~A Hogg}.} \bibinfo{year}{2016}\natexlab{}.
\newblock \bibinfo{booktitle}{\emph{Social identity theory}}.
\newblock \bibinfo{publisher}{Springer}.
\newblock


\bibitem[Huh-Yoo et~al\mbox{.}(2023)]%
        {huh2023help}
\bibfield{author}{\bibinfo{person}{Jina Huh-Yoo}, \bibinfo{person}{Afsaneh Razi}, \bibinfo{person}{Diep~N Nguyen}, \bibinfo{person}{Sampada Regmi}, {and} \bibinfo{person}{Pamela~J Wisniewski}.} \bibinfo{year}{2023}\natexlab{}.
\newblock \showarticletitle{“Help Me:” Examining Youth’s Private Pleas for Support and the Responses Received from Peers via Instagram Direct Messages}. In \bibinfo{booktitle}{\emph{Proceedings of the 2023 CHI Conference on Human Factors in Computing Systems}}. \bibinfo{pages}{1--14}.
\newblock


\bibitem[{Instagram}(2021a)]%
        {instagram_safety_2021}
\bibfield{author}{\bibinfo{person}{{Instagram}}.} \bibinfo{year}{2021}\natexlab{a}.
\newblock \bibinfo{title}{Continuing to Make Instagram Safer for the Youngest Members of Our Community}.
\newblock
\newblock
\urldef\tempurl%
\url{https://about.instagram.com/blog/announcements/continuing-to-make-instagram-safer-for-the-youngest-members-of-our-community}
\showURL{%
\tempurl}
\newblock
\shownote{Accessed: 2025-01-21}.


\bibitem[{Instagram}(2021b)]%
        {instagram_sensitive_content_2021}
\bibfield{author}{\bibinfo{person}{{Instagram}}.} \bibinfo{year}{2021}\natexlab{b}.
\newblock \bibinfo{title}{Introducing Sensitive Content Control}.
\newblock
\newblock
\urldef\tempurl%
\url{https://about.instagram.com/blog/announcements/introducing-sensitive-content-control}
\showURL{%
\tempurl}
\newblock
\shownote{Accessed: 2025-01-21}.


\bibitem[Jung et~al\mbox{.}(2022)]%
        {jung2022social}
\bibfield{author}{\bibinfo{person}{Jaehee Jung}, \bibinfo{person}{David Barron}, \bibinfo{person}{Young-A Lee}, {and} \bibinfo{person}{Viren Swami}.} \bibinfo{year}{2022}\natexlab{}.
\newblock \showarticletitle{Social media usage and body image: Examining the mediating roles of internalization of appearance ideals and social comparisons in young women}.
\newblock \bibinfo{journal}{\emph{Computers in Human Behavior}}  \bibinfo{volume}{135} (\bibinfo{year}{2022}), \bibinfo{pages}{107357}.
\newblock


\bibitem[Kamelabad et~al\mbox{.}(2024)]%
        {kamelabad2024conformity}
\bibfield{author}{\bibinfo{person}{Alireza~M Kamelabad}, \bibinfo{person}{Olov Engwall}, {and} \bibinfo{person}{Gabriel Skantze}.} \bibinfo{year}{2024}\natexlab{}.
\newblock \showarticletitle{Conformity and Trust in Multi-party vs. Individual Human-Robot Interaction}. In \bibinfo{booktitle}{\emph{Proceedings of the 24th ACM International Conference on Intelligent Virtual Agents}}. \bibinfo{pages}{1--9}.
\newblock


\bibitem[Kim and Kim(2023)]%
        {kim2023social}
\bibfield{author}{\bibinfo{person}{Donggyu Kim} {and} \bibinfo{person}{Soomin Kim}.} \bibinfo{year}{2023}\natexlab{}.
\newblock \showarticletitle{Social media affordances of ephemerality and permanence: social comparison, self-esteem, and body image concerns}.
\newblock \bibinfo{journal}{\emph{Social Sciences}} \bibinfo{volume}{12}, \bibinfo{number}{2} (\bibinfo{year}{2023}), \bibinfo{pages}{87}.
\newblock


\bibitem[Kim et~al\mbox{.}(2017)]%
        {kim2017romantic}
\bibfield{author}{\bibinfo{person}{Jung-Eun Kim}, \bibinfo{person}{Emily~C Weinstein}, {and} \bibinfo{person}{Robert~L Selman}.} \bibinfo{year}{2017}\natexlab{}.
\newblock \showarticletitle{Romantic relationship advice from anonymous online helpers: The peer support adolescents exchange}.
\newblock \bibinfo{journal}{\emph{Youth \& Society}} \bibinfo{volume}{49}, \bibinfo{number}{3} (\bibinfo{year}{2017}), \bibinfo{pages}{369--392}.
\newblock


\bibitem[Kruger et~al\mbox{.}(2019)]%
        {kruger2019individual}
\bibfield{author}{\bibinfo{person}{Louis~J Kruger}, \bibinfo{person}{Rachel~F Rodgers}, \bibinfo{person}{Stephanie~J Long}, {and} \bibinfo{person}{Alice~S Lowy}.} \bibinfo{year}{2019}\natexlab{}.
\newblock \showarticletitle{Individual interviews or focus groups? Interview format and women’s self-disclosure}.
\newblock \bibinfo{journal}{\emph{International Journal of Social Research Methodology}} \bibinfo{volume}{22}, \bibinfo{number}{3} (\bibinfo{year}{2019}), \bibinfo{pages}{245--255}.
\newblock


\bibitem[Kruzan et~al\mbox{.}(2021)]%
        {kruzan2021investigating}
\bibfield{author}{\bibinfo{person}{Kaylee~Payne Kruzan}, \bibinfo{person}{Natalya~N Bazarova}, {and} \bibinfo{person}{Janis Whitlock}.} \bibinfo{year}{2021}\natexlab{}.
\newblock \showarticletitle{Investigating self-injury support solicitations and responses on a mobile peer support application}.
\newblock \bibinfo{journal}{\emph{Proceedings of the ACM on human-computer interaction}} \bibinfo{volume}{5}, \bibinfo{number}{CSCW2} (\bibinfo{year}{2021}), \bibinfo{pages}{1--23}.
\newblock


\bibitem[Kwak et~al\mbox{.}(2015)]%
        {kwak2015exploring}
\bibfield{author}{\bibinfo{person}{Haewoon Kwak}, \bibinfo{person}{Jeremy Blackburn}, {and} \bibinfo{person}{Seungyeop Han}.} \bibinfo{year}{2015}\natexlab{}.
\newblock \showarticletitle{Exploring cyberbullying and other toxic behavior in team competition online games}. In \bibinfo{booktitle}{\emph{Proceedings of the 33rd annual ACM conference on human factors in computing systems}}. \bibinfo{pages}{3739--3748}.
\newblock


\bibitem[Lambton-Howard et~al\mbox{.}(2021)]%
        {lambton2021blending}
\bibfield{author}{\bibinfo{person}{Daniel Lambton-Howard}, \bibinfo{person}{Emma Simpson}, \bibinfo{person}{Kim Quimby}, \bibinfo{person}{Ahmed Kharrufa}, \bibinfo{person}{Heidi Hoi Ming~Ng}, \bibinfo{person}{Emma Foster}, {and} \bibinfo{person}{Patrick Olivier}.} \bibinfo{year}{2021}\natexlab{}.
\newblock \showarticletitle{Blending into everyday life: designing a social media-based peer support system}. In \bibinfo{booktitle}{\emph{Proceedings of the 2021 CHI Conference on Human Factors in Computing Systems}}. \bibinfo{pages}{1--14}.
\newblock


\bibitem[Lawler and Nixon(2011)]%
        {lawler2011body}
\bibfield{author}{\bibinfo{person}{Margaret Lawler} {and} \bibinfo{person}{Elizabeth Nixon}.} \bibinfo{year}{2011}\natexlab{}.
\newblock \showarticletitle{Body dissatisfaction among adolescent boys and girls: the effects of body mass, peer appearance culture and internalization of appearance ideals}.
\newblock \bibinfo{journal}{\emph{Journal of youth and adolescence}}  \bibinfo{volume}{40} (\bibinfo{year}{2011}), \bibinfo{pages}{59--71}.
\newblock


\bibitem[Lazuka et~al\mbox{.}(2020)]%
        {Lazuka2020Are}
\bibfield{author}{\bibinfo{person}{Rebecca~F Lazuka}, \bibinfo{person}{Madeline~R Wick}, \bibinfo{person}{P. Keel}, {and} \bibinfo{person}{J. Harriger}.} \bibinfo{year}{2020}\natexlab{}.
\newblock \showarticletitle{Are We There Yet? Progress in Depicting Diverse Images of Beauty in Instagram's Body Positivity Movement.}
\newblock \bibinfo{journal}{\emph{Body image}}  \bibinfo{volume}{34} (\bibinfo{year}{2020}), \bibinfo{pages}{85--93}.
\newblock
\urldef\tempurl%
\url{https://doi.org/10.1016/j.bodyim.2020.05.001}
\showDOI{\tempurl}


\bibitem[Liu and Wei(2018)]%
        {liu2018modeling}
\bibfield{author}{\bibinfo{person}{Bingjie Liu} {and} \bibinfo{person}{Lewen Wei}.} \bibinfo{year}{2018}\natexlab{}.
\newblock \showarticletitle{Modeling social support on social media: Effect of publicness and the underlying mechanisms}.
\newblock \bibinfo{journal}{\emph{Computers in Human Behavior}}  \bibinfo{volume}{87} (\bibinfo{year}{2018}), \bibinfo{pages}{263--275}.
\newblock


\bibitem[Luo and Hancock(2020)]%
        {luo2020self}
\bibfield{author}{\bibinfo{person}{Mufan Luo} {and} \bibinfo{person}{Jeffrey~T Hancock}.} \bibinfo{year}{2020}\natexlab{}.
\newblock \showarticletitle{Self-disclosure and social media: motivations, mechanisms and psychological well-being}.
\newblock \bibinfo{journal}{\emph{Current opinion in psychology}}  \bibinfo{volume}{31} (\bibinfo{year}{2020}), \bibinfo{pages}{110--115}.
\newblock


\bibitem[Mahon and Hevey(2021)]%
        {mahon2021processing}
\bibfield{author}{\bibinfo{person}{Ciara Mahon} {and} \bibinfo{person}{David Hevey}.} \bibinfo{year}{2021}\natexlab{}.
\newblock \showarticletitle{Processing body image on social media: Gender differences in adolescent boys’ and girls’ agency and active coping}.
\newblock \bibinfo{journal}{\emph{Frontiers in psychology}}  \bibinfo{volume}{12} (\bibinfo{year}{2021}), \bibinfo{pages}{626763}.
\newblock


\bibitem[Manikonda and De~Choudhury(2017)]%
        {manikonda2017modeling}
\bibfield{author}{\bibinfo{person}{Lydia Manikonda} {and} \bibinfo{person}{Munmun De~Choudhury}.} \bibinfo{year}{2017}\natexlab{}.
\newblock \showarticletitle{Modeling and understanding visual attributes of mental health disclosures in social media}. In \bibinfo{booktitle}{\emph{Proceedings of the 2017 CHI Conference on Human Factors in Computing Systems}}. \bibinfo{pages}{170--181}.
\newblock


\bibitem[Marwick and Boyd(2014)]%
        {marwick2014networked}
\bibfield{author}{\bibinfo{person}{Alice~E Marwick} {and} \bibinfo{person}{Danah Boyd}.} \bibinfo{year}{2014}\natexlab{}.
\newblock \showarticletitle{Networked privacy: How teenagers negotiate context in social media}.
\newblock \bibinfo{journal}{\emph{New media \& society}} \bibinfo{volume}{16}, \bibinfo{number}{7} (\bibinfo{year}{2014}), \bibinfo{pages}{1051--1067}.
\newblock


\bibitem[Masaki et~al\mbox{.}(2020)]%
        {masaki2020exploring}
\bibfield{author}{\bibinfo{person}{Hiroaki Masaki}, \bibinfo{person}{Kengo Shibata}, \bibinfo{person}{Shui Hoshino}, \bibinfo{person}{Takahiro Ishihama}, \bibinfo{person}{Nagayuki Saito}, {and} \bibinfo{person}{Koji Yatani}.} \bibinfo{year}{2020}\natexlab{}.
\newblock \showarticletitle{Exploring nudge designs to help adolescent SNS users avoid privacy and safety threats}. In \bibinfo{booktitle}{\emph{Proceedings of the 2020 CHI Conference on Human Factors in Computing Systems}}. \bibinfo{pages}{1--11}.
\newblock


\bibitem[Mcdonald et~al\mbox{.}(2024)]%
        {mcdonald2024me}
\bibfield{author}{\bibinfo{person}{Nora Mcdonald}, \bibinfo{person}{John~S Seberger}, {and} \bibinfo{person}{Afsaneh Razi}.} \bibinfo{year}{2024}\natexlab{}.
\newblock \showarticletitle{For Me or Not for Me? The Ease With Which Teens Navigate Accurate and Inaccurate Personalized Social Media Content}. In \bibinfo{booktitle}{\emph{Proceedings of the CHI Conference on Human Factors in Computing Systems}}. \bibinfo{pages}{1--7}.
\newblock


\bibitem[McGregor and Li(2019)]%
        {mcgregor201973}
\bibfield{author}{\bibinfo{person}{Kyle~Aaron McGregor} {and} \bibinfo{person}{Joanna Li}.} \bibinfo{year}{2019}\natexlab{}.
\newblock \showarticletitle{73. Fake Instagrams for real conversation: A thematic analysis of the hidden social media life of teenagers}.
\newblock \bibinfo{journal}{\emph{Journal of Adolescent Health}} \bibinfo{volume}{64}, \bibinfo{number}{2} (\bibinfo{year}{2019}), \bibinfo{pages}{S39--S40}.
\newblock


\bibitem[McNally et~al\mbox{.}(2018)]%
        {mcnally2018co}
\bibfield{author}{\bibinfo{person}{Brenna McNally}, \bibinfo{person}{Priya Kumar}, \bibinfo{person}{Chelsea Hordatt}, \bibinfo{person}{Matthew~Louis Mauriello}, \bibinfo{person}{Shalmali Naik}, \bibinfo{person}{Leyla Norooz}, \bibinfo{person}{Alazandra Shorter}, \bibinfo{person}{Evan Golub}, {and} \bibinfo{person}{Allison Druin}.} \bibinfo{year}{2018}\natexlab{}.
\newblock \showarticletitle{Co-designing mobile online safety applications with children}. In \bibinfo{booktitle}{\emph{Proceedings of the 2018 CHI Conference on Human Factors in Computing Systems}}. \bibinfo{pages}{1--9}.
\newblock


\bibitem[Mills and Fuller-Tyszkiewicz(2017)]%
        {mills2017fat}
\bibfield{author}{\bibinfo{person}{Jacqueline Mills} {and} \bibinfo{person}{Matthew Fuller-Tyszkiewicz}.} \bibinfo{year}{2017}\natexlab{}.
\newblock \showarticletitle{Fat talk and body image disturbance: A systematic review and meta-analysis}.
\newblock \bibinfo{journal}{\emph{Psychology of Women Quarterly}} \bibinfo{volume}{41}, \bibinfo{number}{1} (\bibinfo{year}{2017}), \bibinfo{pages}{114--129}.
\newblock


\bibitem[Mills et~al\mbox{.}(2019)]%
        {mills2019impact}
\bibfield{author}{\bibinfo{person}{Jacqueline Mills}, \bibinfo{person}{Olivia Mort}, {and} \bibinfo{person}{Steven Trawley}.} \bibinfo{year}{2019}\natexlab{}.
\newblock \showarticletitle{The impact of different responses to fat talk on body image and socioemotional outcomes}.
\newblock \bibinfo{journal}{\emph{Body image}}  \bibinfo{volume}{29} (\bibinfo{year}{2019}), \bibinfo{pages}{149--155}.
\newblock


\bibitem[Milton et~al\mbox{.}(2023)]%
        {milton2023see}
\bibfield{author}{\bibinfo{person}{Ashlee Milton}, \bibinfo{person}{Leah Ajmani}, \bibinfo{person}{Michael~Ann DeVito}, {and} \bibinfo{person}{Stevie Chancellor}.} \bibinfo{year}{2023}\natexlab{}.
\newblock \showarticletitle{“I See Me Here”: Mental Health Content, Community, and Algorithmic Curation on TikTok}. In \bibinfo{booktitle}{\emph{Proceedings of the 2023 CHI conference on human factors in computing systems}}. \bibinfo{pages}{1--17}.
\newblock


\bibitem[Mostafavi(2022)]%
        {Mostafavi_2022}
\bibfield{author}{\bibinfo{person}{Beata Mostafavi}.} \bibinfo{year}{2022}\natexlab{}.
\newblock \bibinfo{title}{Fighting negative body image issues in kids and teens}.
\newblock
\newblock
\urldef\tempurl%
\url{https://www.michiganmedicine.org/health-lab/fighting-negative-body-image-issues-kids-and-teens}
\showURL{%
\tempurl}


\bibitem[Naslund et~al\mbox{.}(2014)]%
        {naslund2014naturally}
\bibfield{author}{\bibinfo{person}{John~A Naslund}, \bibinfo{person}{Stuart~W Grande}, \bibinfo{person}{Kelly~A Aschbrenner}, {and} \bibinfo{person}{Glyn Elwyn}.} \bibinfo{year}{2014}\natexlab{}.
\newblock \showarticletitle{Naturally occurring peer support through social media: the experiences of individuals with severe mental illness using YouTube}.
\newblock \bibinfo{journal}{\emph{PLOS one}} \bibinfo{volume}{9}, \bibinfo{number}{10} (\bibinfo{year}{2014}), \bibinfo{pages}{e110171}.
\newblock


\bibitem[Nova et~al\mbox{.}(2022)]%
        {nova2022cultivating}
\bibfield{author}{\bibinfo{person}{Fayika~Farhat Nova}, \bibinfo{person}{Amanda Coupe}, \bibinfo{person}{Elizabeth~D Mynatt}, \bibinfo{person}{Shion Guha}, {and} \bibinfo{person}{Jessica~A Pater}.} \bibinfo{year}{2022}\natexlab{}.
\newblock \showarticletitle{Cultivating the community: inferring influence within eating disorder networks on twitter}.
\newblock \bibinfo{journal}{\emph{Proceedings of the ACM on Human-Computer Interaction}} \bibinfo{volume}{6}, \bibinfo{number}{GROUP} (\bibinfo{year}{2022}), \bibinfo{pages}{1--33}.
\newblock


\bibitem[Oguine et~al\mbox{.}(2024)]%
        {oguine2024internet}
\bibfield{author}{\bibinfo{person}{Ozioma~C Oguine}, \bibinfo{person}{Jinkyung~Katie Park}, \bibinfo{person}{Mamtaj Akter}, \bibinfo{person}{Johanna Olesk}, \bibinfo{person}{Abdulmalik Alluhidan}, \bibinfo{person}{Pamela Wisniewski}, {and} \bibinfo{person}{Karla Badillo-Urquiola}.} \bibinfo{year}{2024}\natexlab{}.
\newblock \showarticletitle{How the Internet Facilitates Adverse Childhood Experiences for Youth Who Self-Identify as in Need of Services}.
\newblock \bibinfo{journal}{\emph{arXiv preprint arXiv:2410.16507}} (\bibinfo{year}{2024}).
\newblock


\bibitem[Orlando(2020)]%
        {orlando2020peer}
\bibfield{author}{\bibinfo{person}{Carissa~M Orlando}.} \bibinfo{year}{2020}\natexlab{}.
\newblock \emph{\bibinfo{title}{Peer Intervention with Suicidal Disclosures on Social Media: Does the Bystander Effect Play a Role?}}
\newblock \bibinfo{thesistype}{Ph.\,D. Dissertation}. \bibinfo{school}{University of South Carolina}.
\newblock


\bibitem[Park et~al\mbox{.}(2023)]%
        {park2023affordances}
\bibfield{author}{\bibinfo{person}{Jinkyung Park}, \bibinfo{person}{Irina Lediaeva}, \bibinfo{person}{Maria Lopez}, \bibinfo{person}{Amy Godfrey}, \bibinfo{person}{Kapil~Chalil Madathil}, \bibinfo{person}{Heidi Zinzow}, {and} \bibinfo{person}{Pamela Wisniewski}.} \bibinfo{year}{2023}\natexlab{}.
\newblock \showarticletitle{How affordances and social norms shape the discussion of harmful social media challenges on reddit}.
\newblock \bibinfo{journal}{\emph{Human Factors in Healthcare}}  \bibinfo{volume}{3} (\bibinfo{year}{2023}), \bibinfo{pages}{100042}.
\newblock


\bibitem[Park et~al\mbox{.}(2024)]%
        {park2024resilience}
\bibfield{author}{\bibinfo{person}{Jinkyung~Katie Park}, \bibinfo{person}{Mamtaj Akter}, \bibinfo{person}{Pamela Wisniewski}, {and} \bibinfo{person}{Karla Badillo-Urquiola}.} \bibinfo{year}{2024}\natexlab{}.
\newblock \showarticletitle{It’s Still Complicated: From Privacy-Invasive Parental Control to Teen-Centric Solutions for Digital Resilience}.
\newblock \bibinfo{journal}{\emph{IEEE Security \& Privacy}} \bibinfo{volume}{22}, \bibinfo{number}{5} (\bibinfo{year}{2024}), \bibinfo{pages}{52--62}.
\newblock
\urldef\tempurl%
\url{https://doi.org/10.1109/MSEC.2024.3417804}
\showDOI{\tempurl}


\bibitem[Pater et~al\mbox{.}(2019a)]%
        {pater2019exploring}
\bibfield{author}{\bibinfo{person}{Jessica~A Pater}, \bibinfo{person}{Brooke Farrington}, \bibinfo{person}{Alycia Brown}, \bibinfo{person}{Lauren~E Reining}, \bibinfo{person}{Tammy Toscos}, {and} \bibinfo{person}{Elizabeth~D Mynatt}.} \bibinfo{year}{2019}\natexlab{a}.
\newblock \showarticletitle{Exploring indicators of digital self-harm with eating disorder patients: A case study}.
\newblock \bibinfo{journal}{\emph{Proceedings of the ACM on human-computer interaction}} \bibinfo{volume}{3}, \bibinfo{number}{CSCW} (\bibinfo{year}{2019}), \bibinfo{pages}{1--26}.
\newblock


\bibitem[Pater et~al\mbox{.}(2019b)]%
        {pater2019notjustgirls}
\bibfield{author}{\bibinfo{person}{Jessica~A Pater}, \bibinfo{person}{Lauren~E Reining}, \bibinfo{person}{Andrew~D Miller}, \bibinfo{person}{Tammy Toscos}, {and} \bibinfo{person}{Elizabeth~D Mynatt}.} \bibinfo{year}{2019}\natexlab{b}.
\newblock \showarticletitle{" Notjustgirls" Exploring Male-related Eating Disordered Content across Social Media Platforms}. In \bibinfo{booktitle}{\emph{Proceedings of the 2019 CHI Conference on Human Factors in Computing Systems}}. \bibinfo{pages}{1--13}.
\newblock


\bibitem[Pedalino and Camerini(2022)]%
        {pedalino2022instagram}
\bibfield{author}{\bibinfo{person}{Federica Pedalino} {and} \bibinfo{person}{Anne-Linda Camerini}.} \bibinfo{year}{2022}\natexlab{}.
\newblock \showarticletitle{Instagram use and body dissatisfaction: The mediating role of upward social comparison with peers and influencers among young females}.
\newblock \bibinfo{journal}{\emph{International journal of environmental research and public health}} \bibinfo{volume}{19}, \bibinfo{number}{3} (\bibinfo{year}{2022}), \bibinfo{pages}{1543}.
\newblock


\bibitem[P{\'e}rez-Torres(2024)]%
        {perez2024social}
\bibfield{author}{\bibinfo{person}{Vanesa P{\'e}rez-Torres}.} \bibinfo{year}{2024}\natexlab{}.
\newblock \showarticletitle{Social media: a digital social mirror for identity development during adolescence}.
\newblock \bibinfo{journal}{\emph{Current Psychology}} (\bibinfo{year}{2024}), \bibinfo{pages}{1--11}.
\newblock


\bibitem[Razi et~al\mbox{.}(2022)]%
        {razi2022instagram}
\bibfield{author}{\bibinfo{person}{Afsaneh Razi}, \bibinfo{person}{Ashwaq AlSoubai}, \bibinfo{person}{Seunghyun Kim}, \bibinfo{person}{Nurun Naher}, \bibinfo{person}{Shiza Ali}, \bibinfo{person}{Gianluca Stringhini}, \bibinfo{person}{Munmun De~Choudhury}, {and} \bibinfo{person}{Pamela~J Wisniewski}.} \bibinfo{year}{2022}\natexlab{}.
\newblock \showarticletitle{Instagram Data Donation: A Case Study on Collecting Ecologically Valid Social Media Data for the Purpose of Adolescent Online Risk Detection}. In \bibinfo{booktitle}{\emph{CHI Conference on Human Factors in Computing Systems Extended Abstracts}}. \bibinfo{pages}{1--9}.
\newblock


\bibitem[Razi et~al\mbox{.}(2020)]%
        {razi2020let}
\bibfield{author}{\bibinfo{person}{Afsaneh Razi}, \bibinfo{person}{Karla Badillo-Urquiola}, {and} \bibinfo{person}{Pamela~J Wisniewski}.} \bibinfo{year}{2020}\natexlab{}.
\newblock \showarticletitle{Let's talk about sext: How adolescents seek support and advice about their online sexual experiences}. In \bibinfo{booktitle}{\emph{Proceedings of the 2020 CHI Conference on Human Factors in Computing Systems}}. \bibinfo{pages}{1--13}.
\newblock


\bibitem[Ronzhyn et~al\mbox{.}(2023)]%
        {ronzhyn2023defining}
\bibfield{author}{\bibinfo{person}{Alexander Ronzhyn}, \bibinfo{person}{Ana~Sofia Cardenal}, {and} \bibinfo{person}{Albert Batlle~Rubio}.} \bibinfo{year}{2023}\natexlab{}.
\newblock \showarticletitle{Defining affordances in social media research: A literature review}.
\newblock \bibinfo{journal}{\emph{New Media \& Society}} \bibinfo{volume}{25}, \bibinfo{number}{11} (\bibinfo{year}{2023}), \bibinfo{pages}{3165--3188}.
\newblock


\bibitem[Rubio-Hurtado et~al\mbox{.}(2022)]%
        {rubio2022youths}
\bibfield{author}{\bibinfo{person}{Mar{\'\i}a-Jos{\'e} Rubio-Hurtado}, \bibinfo{person}{Marc Fuertes-Alpiste}, \bibinfo{person}{Francesc Mart{\'\i}nez-Olmo}, {and} \bibinfo{person}{Jordi Quintana}.} \bibinfo{year}{2022}\natexlab{}.
\newblock \showarticletitle{Youths' Posting Practices on Social Media for Digital Storytelling.}
\newblock \bibinfo{journal}{\emph{Journal of new approaches in educational research}} \bibinfo{volume}{11}, \bibinfo{number}{1} (\bibinfo{year}{2022}), \bibinfo{pages}{97--113}.
\newblock


\bibitem[Sagrera et~al\mbox{.}(2022)]%
        {sagrera2022social}
\bibfield{author}{\bibinfo{person}{Caroline~E Sagrera}, \bibinfo{person}{Johnette Magner}, \bibinfo{person}{Jazzlynn Temple}, \bibinfo{person}{Robert Lawrence}, \bibinfo{person}{Timothy~J Magner}, \bibinfo{person}{Victor~J Avila-Quintero}, \bibinfo{person}{Pamela McPherson}, \bibinfo{person}{Laura~Lane Alderman}, \bibinfo{person}{Mohammad Alfrad~Nobel Bhuiyan}, \bibinfo{person}{James~C Patterson}, {et~al\mbox{.}}} \bibinfo{year}{2022}\natexlab{}.
\newblock \showarticletitle{Social media use and body image issues among adolescents in a vulnerable Louisiana community}.
\newblock \bibinfo{journal}{\emph{Frontiers in Psychiatry}}  \bibinfo{volume}{13} (\bibinfo{year}{2022}), \bibinfo{pages}{1001336}.
\newblock


\bibitem[Saiphoo and Vahedi(2019)]%
        {saiphoo2019meta}
\bibfield{author}{\bibinfo{person}{Alyssa~N Saiphoo} {and} \bibinfo{person}{Zahra Vahedi}.} \bibinfo{year}{2019}\natexlab{}.
\newblock \showarticletitle{A meta-analytic review of the relationship between social media use and body image disturbance}.
\newblock \bibinfo{journal}{\emph{Computers in human behavior}}  \bibinfo{volume}{101} (\bibinfo{year}{2019}), \bibinfo{pages}{259--275}.
\newblock


\bibitem[Sampson et~al\mbox{.}(2020)]%
        {sampson2020effect}
\bibfield{author}{\bibinfo{person}{Ariane Sampson}, \bibinfo{person}{Huw~G Jeremiah}, \bibinfo{person}{Manoharan Andiappan}, {and} \bibinfo{person}{J~Tim Newton}.} \bibinfo{year}{2020}\natexlab{}.
\newblock \showarticletitle{The effect of viewing idealised smile images versus nature images via social media on immediate facial satisfaction in young adults: A randomised controlled trial}.
\newblock \bibinfo{journal}{\emph{Journal of orthodontics}} \bibinfo{volume}{47}, \bibinfo{number}{1} (\bibinfo{year}{2020}), \bibinfo{pages}{55--64}.
\newblock


\bibitem[Sharma et~al\mbox{.}(2020)]%
        {sharma2020computational}
\bibfield{author}{\bibinfo{person}{Ashish Sharma}, \bibinfo{person}{Adam~S Miner}, \bibinfo{person}{David~C Atkins}, {and} \bibinfo{person}{Tim Althoff}.} \bibinfo{year}{2020}\natexlab{}.
\newblock \showarticletitle{A computational approach to understanding empathy expressed in text-based mental health support}.
\newblock \bibinfo{journal}{\emph{arXiv preprint arXiv:2009.08441}} (\bibinfo{year}{2020}).
\newblock


\bibitem[Sharpe(2015)]%
        {sharpe2015chi}
\bibfield{author}{\bibinfo{person}{Donald Sharpe}.} \bibinfo{year}{2015}\natexlab{}.
\newblock \showarticletitle{Chi-square test is statistically significant: Now what?}
\newblock \bibinfo{journal}{\emph{Practical Assessment, Research, and Evaluation}} \bibinfo{number}{1} (\bibinfo{year}{2015}), \bibinfo{pages}{8}.
\newblock


\bibitem[Solano and Dunnam(1985)]%
        {solano1985two}
\bibfield{author}{\bibinfo{person}{Cecilia~H Solano} {and} \bibinfo{person}{Mina Dunnam}.} \bibinfo{year}{1985}\natexlab{}.
\newblock \showarticletitle{Two's company: Self-disclosure and reciprocity in triads versus dyads}.
\newblock \bibinfo{journal}{\emph{Social Psychology Quarterly}} (\bibinfo{year}{1985}), \bibinfo{pages}{183--187}.
\newblock


\bibitem[Syed et~al\mbox{.}(2024)]%
        {syed2024machine}
\bibfield{author}{\bibinfo{person}{Sara Syed}, \bibinfo{person}{Zainab Iftikhar}, \bibinfo{person}{Amy~Wei Xiao}, {and} \bibinfo{person}{Jeff Huang}.} \bibinfo{year}{2024}\natexlab{}.
\newblock \showarticletitle{Machine and Human Understanding of Empathy in Online Peer Support: A Cognitive Behavioral Approach}. In \bibinfo{booktitle}{\emph{Proceedings of the CHI Conference on Human Factors in Computing Systems}}. \bibinfo{pages}{1--13}.
\newblock


\bibitem[Taylor et~al\mbox{.}(1979)]%
        {taylor1979sharing}
\bibfield{author}{\bibinfo{person}{Ralph~B Taylor}, \bibinfo{person}{Clinton~B De~Soto}, {and} \bibinfo{person}{Robert Lieb}.} \bibinfo{year}{1979}\natexlab{}.
\newblock \showarticletitle{Sharing secrets: Disclosure and discretion in dyads and triads.}
\newblock \bibinfo{journal}{\emph{Journal of Personality and Social Psychology}} \bibinfo{volume}{37}, \bibinfo{number}{7} (\bibinfo{year}{1979}), \bibinfo{pages}{1196}.
\newblock


\bibitem[Thorn et~al\mbox{.}(2023)]%
        {thorn2023motivations}
\bibfield{author}{\bibinfo{person}{Pinar Thorn}, \bibinfo{person}{Louise La~Sala}, \bibinfo{person}{Sarah Hetrick}, \bibinfo{person}{Simon Rice}, \bibinfo{person}{Michelle Lamblin}, {and} \bibinfo{person}{Jo Robinson}.} \bibinfo{year}{2023}\natexlab{}.
\newblock \showarticletitle{Motivations and perceived harms and benefits of online communication about self-harm: An interview study with young people}.
\newblock \bibinfo{journal}{\emph{Digital health}}  \bibinfo{volume}{9} (\bibinfo{year}{2023}), \bibinfo{pages}{20552076231176689}.
\newblock


\bibitem[Toenders et~al\mbox{.}(2024)]%
        {toenders2024developing}
\bibfield{author}{\bibinfo{person}{Yara~J Toenders}, \bibinfo{person}{Hannah Dorsman}, \bibinfo{person}{Renske van~der Cruijsen}, {and} \bibinfo{person}{Eveline~A Crone}.} \bibinfo{year}{2024}\natexlab{}.
\newblock \showarticletitle{Developing body estimation in adolescence is associated with neural regions that support self-concept}.
\newblock \bibinfo{journal}{\emph{Social Cognitive and Affective Neuroscience}} (\bibinfo{year}{2024}), \bibinfo{pages}{nsae042}.
\newblock


\bibitem[Vandenbosch et~al\mbox{.}(2022)]%
        {vandenbosch2022social}
\bibfield{author}{\bibinfo{person}{Laura Vandenbosch}, \bibinfo{person}{Jasmine Fardouly}, {and} \bibinfo{person}{Marika Tiggemann}.} \bibinfo{year}{2022}\natexlab{}.
\newblock \showarticletitle{Social media and body image: Recent trends and future directions}.
\newblock \bibinfo{journal}{\emph{Current opinion in psychology}}  \bibinfo{volume}{45} (\bibinfo{year}{2022}), \bibinfo{pages}{101289}.
\newblock


\bibitem[Wang et~al\mbox{.}(2022)]%
        {wang2022tiktok}
\bibfield{author}{\bibinfo{person}{Chun-Han Wang}, \bibinfo{person}{Stephen Tsung-Han Sher}, \bibinfo{person}{Isabela Salman}, \bibinfo{person}{Kelly Janek}, {and} \bibinfo{person}{Chia-Fang Chung}.} \bibinfo{year}{2022}\natexlab{}.
\newblock \showarticletitle{“TikTok made me do it”: teenagers’ perception and use of food content on TikTok}. In \bibinfo{booktitle}{\emph{Proceedings of the 21st annual ACM interaction design and children conference}}. \bibinfo{pages}{458--463}.
\newblock


\bibitem[Wisniewski et~al\mbox{.}(2016)]%
        {wisniewski2016dear}
\bibfield{author}{\bibinfo{person}{Pamela Wisniewski}, \bibinfo{person}{Heng Xu}, \bibinfo{person}{Mary~Beth Rosson}, \bibinfo{person}{Daniel~F Perkins}, {and} \bibinfo{person}{John~M Carroll}.} \bibinfo{year}{2016}\natexlab{}.
\newblock \showarticletitle{Dear diary: Teens reflect on their weekly online risk experiences}. In \bibinfo{booktitle}{\emph{Proceedings of the 2016 CHI Conference on Human Factors in Computing Systems}}. \bibinfo{pages}{3919--3930}.
\newblock


\bibitem[Yang et~al\mbox{.}(2022)]%
        {yang2022longitudinal}
\bibfield{author}{\bibinfo{person}{Jiping Yang}, \bibinfo{person}{Shuang Li}, \bibinfo{person}{Ling Gao}, {and} \bibinfo{person}{Xingchao Wang}.} \bibinfo{year}{2022}\natexlab{}.
\newblock \showarticletitle{Longitudinal associations among peer pressure, moral disengagement and cyberbullying perpetration in adolescents}.
\newblock \bibinfo{journal}{\emph{Computers in Human Behavior}}  \bibinfo{volume}{137} (\bibinfo{year}{2022}), \bibinfo{pages}{107420}.
\newblock


\bibitem[You and Lee(2019)]%
        {you2019bystander}
\bibfield{author}{\bibinfo{person}{Leping You} {and} \bibinfo{person}{Yu-Hao Lee}.} \bibinfo{year}{2019}\natexlab{}.
\newblock \showarticletitle{The bystander effect in cyberbullying on social network sites: Anonymity, group size, and intervention intentions}.
\newblock \bibinfo{journal}{\emph{Telematics and Informatics}}  \bibinfo{volume}{45} (\bibinfo{year}{2019}), \bibinfo{pages}{101284}.
\newblock


\end{thebibliography}
\appendix

\end{document}